\documentclass[preprint,12pt]{aastex}
\newcommand{\degree}{\mbox{$^{\circ}$}}
\newcommand{\am}{\mbox{\arcmin}}
\newcommand{\as}{\mbox{\arcsec}}

\newcommand{\kms}{\mbox{km s$^{-1}$}}% km/s

\def\lsim {$\rlap{\raise.4ex\hbox{$<$}}\lower.55ex\hbox{$\sim$}\,$}

\newcommand{\lsun}{\mbox{L$_\odot$}}% Lsun
\newcommand{\msun}{\mbox{M$_\odot$}}% Msun

\input{epsf}

\def\deg{{$^\circ$}}
\begin{document}

%%%%%%%%%%%%%%%%%% title %%%%%%%%%%%%%%%%%%%%%%%%%%%%%%%%%%%%%%%%

\title{\bf Molecular Line Observations of the Small Protostellar Group L1251B}
\author {Jeong-Eun Lee\altaffilmark{1,2},
James Di Francesco\altaffilmark{3},
Tyler L. Bourke\altaffilmark{4},
Neal J. Evans II\altaffilmark{5},
Jingwen Wu\altaffilmark{4},
}
\altaffiltext{1}{Department of Astronomy and Space Science, Sejong University, 
Seoul 143-747, Korea; jelee@sejong.ac.kr} 
\altaffiltext{2}{Hubble Fellow, Physics and Astronomy Department, The University
of California at Los Angeles, PAB, 430 Portola Plaza, Box 951547, Los Angeles, 
CA 90095-1547}
\altaffiltext{3}{National Research Council of Canada, Herzberg Institute of 
Astrophysics, 5071 West Saanich Road, Victoria, BC V9E 2E7, Canada; 
james.difrancesco@nrc-cnrc.gc.ca}
\altaffiltext{4}{Smithsonian Astrophysical Observatory, 60 Garden Street, 
Cambridge, MA 02138; tbourke@cfa.harvard.edu}
\altaffiltext{5}{Astronomy Department, The University of Texas at Austin,
1 University Station C1400, Austin, TX 78712-0259; 
nje@astro.as.utexas.edu, jingwen@astro.as.utexas.edu}

%%%%%%%%%%%%%%%%%% abstract %%%%%%%%%%%%%%%%%%%%%%%%%%%%%%%%%%%%%%%%
 
\begin{abstract}
We present molecular line observations of L1251B, a small group of pre- and
protostellar objects, and its immediate environment in the dense C$^{18}$O
core L1251E.  These data are complementary to near-infrared, submillimeter
and millimeter continuum observations reported by Lee et al. (2006, ApJ, 648,
491; Paper I).  The single-dish data of L1251B described here show very
complex kinematics including infall, rotation and outflow motions, and
the interferometer data reveal these in greater detail.
Interferometer data of N$_{2}$H$^{+}$ 1$-$0 suggest a very rapidly rotating
flattened envelope between two young stellar objects, IRS1 and IRS2.
Also, interferometer data of CO 2$-$1 
resolve the outflow associated with L1251B seen in single-dish maps
into a few narrow and compact components.  Furthermore, the
high resolution data support recent theoretical studies of molecular
depletions and enhancements that accompany the formation of protostars
within dense cores.  Beyond L1251B, single-dish data are also presented of
a dense core located $\sim$150\as\ to the east that, in Paper I, was detected
at 850 \micron\ but has no associated point sources at near- and mid-infrared
wavelengths.  
The relative brightness between molecules, which have different chemical 
timescales, suggests it is less chemically evolved than L1251B.  This core may
be a site for future star formation, however, since line profiles of HCO$^{+}$,
CS, and HCN show asymmetry with a stronger blue peak, which is 
interpreted as an infall signature.
\end{abstract}


\section{INTRODUCTION}

Most young stars in our Galaxy formed within groups or clusters (Lada and Lada
2003).  Small, nearby groups of protostellar objects provide compelling cases 
for investigating the physical processes involved with such formation since the 
numbers involved are relatively small (e.g., $N$ $<$ 10) and the interactions 
of these objects with themselves and their environments are less complicated.
For example, observational study of kinematics associated with young stars 
within groups and their immediate dense gas surroundings could provide evidence 
for a specific formation scenario, e.g., fragmentation or triggering, that is 
easier to see because of the simpler nature of groups relative to clusters.

Kinematic study of the gas surroundings of protostellar objects is achieved
through observation and interpretation of molecular rotational transitions,
since line profiles can be shaped by the motions within the source gas (see
Di Francesco et al. 2006 for a review.)  For example, outflow motions from
young stellar objects were first seen in the low-intensity 
wings of CO 1$-$0 emission (e.g., Snell, Loren \& Plambeck 1980).  In addition,
rotational motions in dense cores have been suggested from the detection of
the shift of centroid velocities along different lines of sight in optically 
thin molecular transitions (e.g,. Goodman et al. 1993).  
Finally, infall motions in dense cores have 
been the interpretation of profiles of moderately optically thick lines that 
are asymmetric with a stronger blue peak relative to optically thin lines 
(e.g., Zhou et al. 1993; Myers, Evans \& Ohashi 2000).  
Such kinematic evidence has come primarily 
from single-dish telescopes, i.e., of relatively low spatial resolution, in 
studies of isolated protostellar objects.  For protostellar groups, however, 
the surface density of objects can be high, and single-dish telescopes may not 
have the resolution needed to associate kinematics with particular protostars 
within the group.  Instead, data from interferometers are very highly suited 
for such studies, especially when used in tandem with data from single-dish 
telescopes.

The use of molecular lines to study kinematics is complicated by potentially
significant molecular abundance variations along the lines of sight toward 
dense gas both before and after star formation.  Such abundance changes can
be due to chemical evolution, intertwined with the YSO luminosity evolution
since luminosity affects the temperatures of nearby dust grains that interact 
with gaseous molecules (see Rawlings \& Yates 2001; Doty et al. 2002; Lee, 
Bergin \& Evans 2004).  Of course, YSO luminosity evolution is itself 
intertwined with dynamical evolution of the surrounding gas, as much of the 
luminosity originates from accretion.  Hence, a holistic approach in general 
is required that includes luminosity, dynamics and chemistry, to understand 
best the environments from which stars form (see Bergin et al. 1997 and Aikawa 
et al. 2001 for early examples of such models).

Located at 300 pc $\pm$ 50 pc (Kun \& Prusti 1993), L1251B is an excellent 
example of a small, nearby group of protostellar objects where molecular line 
observations can probe the kinematics and chemistry of its environment and 
yield clues about its origins.  L1251B is located within L1251E, the densest
C$^{18}$O core in L1251 (Sato et al. 1994).  L1251B is associated with a 
single IRAS point source, IRAS 22376+7455, and was found associated with CO 
outflow by Sato \& Fukui (1989).  In Lee et al. (2006; Paper I), we showed 
via single-dish data that IRAS 22376+7455 was associated with a local maximum 
of submillimeter emission, although significant peaks of 850 $\mu$m emission 
were also seen $\sim$150\as\ to the east.  In addition, we showed Spitzer 
Space Telescope (SST) observations of the region that revealed $\sim$20 YSOs 
within $\sim$1 pc$^2$ covering L1251B and its surroundings, indicative of the 
formation of a group of low-mass stars.  Within L1251B itself, these data 
revealed three Class 0/I objects (IRS1, IRS2, and IRS4) and three Class II 
objects (IRS3, IRS5, and IRS6), within a projected diameter of 55\as.  (No 
near- or mid-infrared sources were found to be associated with other locations 
of submillimeter emission adjacent to L1251B.)  Finally, we showed 
Submillimeter Array (SMA) continuum observations of L1251B in Paper I that 
revealed two compact locations of high column density between IRS1 and IRS2 
that may be starless condensations.  Paper I revealed clearly L1251B as a 
region of recent (and possibly continuing) star formation within a small 
group.

In this paper, we examine line emission associated with L1251B and the core 
$\sim$150\as\ to its east (henceforth, the ``east core"), to probe kinematic
and chemical processes in the region.  The three embedded Class 0/I sources 
and two starless objects within L1251B are especially emphasized; IRS1 is the 
most luminous object, IRS2 is located southeast of IRS1 and is associated with 
a near-infrared bipolar nebula, and IRS4 is located at the west of IRS1.  
The two starless objects are named as SMA-N (the northern one) and SMA-S 
(the southern one) in this paper.  
Besides the outflow noted by Sato 
\& Fukui, earlier molecular line observations have also provided evidence for
rotational and infall motions in L1251B.  For example, a velocity gradient of 
$\sim$2.3 km s$^{-1}$ pc$^{-1}$ in the northeast--southwest direction across 
L1251B suggestive of rotational motion was detected by Goodman et al. (1993) 
in NH$_{3}$ (1,1) (also Sato et al. 1994; T$\acute{\rm o}$th \& Walmsley 1996; 
Anglada et al.  1997; Morata et al. 1997).  In addition, infall motions toward 
L1251B were suggested from the detections of  asymmetric profiles with 
stronger blue peaks in
HCO$^+$ 1$-$0 and 3$-$2, CS 2$-$1 and H$_2$CO $2_{12}$--$1_{11}$ 
by Gregersen et al. (2000) and Mardones et al. (1997), though observations were 
published only 
toward the IRAS point source position.  Since the relative intensity of a given 
molecular line depends upon distributions of density, temperature and molecular 
abundance along the line of sight, many other transitions and positions needed 
to be observed within L1251B and its surroundings to understand correctly their 
physical conditions and dynamical processes.  Here, we have gathered with 
single-dish telescopes or interferometers new observations of L1251B in CS 
2$-$1, 3$-$2, and 5$-$4, HCN 1$-$0, CO 2$-$1, $^{13}$CO 2$-$1, C$^{18}$O 2$-$1, 
HCO$^+$ 1$-$0 and 3$-$2, H$^{13}$CO$^+$ 1$-$0, DCO$^+$ 3$-$2, N$_2$H$^+$ 1$-$0 
and H$_2$CO $3_{12}$--$2_{11}$.

Details of the acquisition of observations of L1251B for this paper are 
summarized in \S 2.  Results and analysis of the data are described in \S 
3 and \S 4.  In \S 5, we discuss the chemical and dynamical conditions in 
L1251B and the east core.  
Finally, a summary of the conclusions of this paper appears in \S 6.

\section{OBSERVATIONS}

\subsection{The Five College Radio Astronomy Observatory (FCRAO) Observations}

Observations of N$_2$H$^+$ 1$-$0 toward L1251B were made in December 1996 with 
the FCRAO 14-m telescope near New Salem, MA using QUARRY, the 15-element focal 
plane heterodyne receiver array.  Observations of HCO$^+$ 1$-$0 and 
H$^{13}$CO$^+$ 1$-$0 were made toward L1251B also at FCRAO in April 1997 and 
May 2000, using the QUARRY (1997) and the 32-element SEQUOIA (2000) 
focal plane arrays respectively.  The autocorrelator spectrometer was 
configured with a band width of 20 MHz over 1024 channels, providing a channel 
separation of $\sim$20 kHz ($\sim$0.07 \kms).  Typical system temperatures 
were 500-700 K (QUARRY) and 200-250 K (SEQUOIA).  Observations of HCN J=1$-$0 
and CS 2$-$1 toward L1251B were also made at FCRAO in February 2005 in OTF 
mode using SEQUOIA with a system temperature of $\sim$120 K.  A velocity 
resolution of 0.1 km s$^{-1}$ was achieved with the 25 MHz band width on the 
dual channel correlator and the data were acquired in frequency-switching mode 
with an 8 MHz throw.  The average pointing accuracy was about 4.5\arcsec.
Table 1 summarizes observational details of the FCRAO data, including the 
frequency ($\nu$), velocity resolution ($\delta v$), the FWHM beam size 
($\theta_{mb}$), the main beam efficiency ($\eta_{\rm mb}$), and the 
observing date for each line.

\subsection{The Caltech Submillimeter Observatory (CSO) Observations}

We observed the HCO$^+$ 3$-$2, CO 2$-$1, and DCO$^+$ 3$-$2 lines with the 
10.4 m telescope of the CSO at Mauna Kea, HI in June 1997 and July 2003.
HCO$^+$ 3$-$2 had been observed as part of the survey for infall signatures 
by Gregersen et al. (2000).  We used an SIS receiver with an acousto-optic
spectrometer (AOS) with a 50 MHz band width and 1024 channels.  The frequency 
resolution was about 2 to 2.5 channels, i.e., $\sim$0.13 $\rm km~s ^{-1}$ at 
220 GHz. The pointing uncertainty was approximately 4$\arcsec$ on average.  
Table 1 also summarizes observational details of these data.

\subsection{The Caltech Owens Valley Radio Observatory (OVRO) Millimeter 
Array Observations}

L1251B was observed in several transitions and continuum emission in the 1 
mm or 3 mm bands of the OVRO Millimeter Array near Big Pine, CA over several 
observing seasons.  The continuum data have been presented in Paper I.  Table 
2 summarizes the line data, i.e., the frequencies observed and the band widths 
used, as well as the velocity channel spacings, the synthesized beam FWHMs, 
and the $1\sigma$ rms values achieved.  

In Fall 1997 and Spring 1998, H$_{2}$CO J$_{K_{-1}K_{+1}}$ = 3$_{12}$--2$_{11}$
at 225.6977750 GHz (Pickett et al. 1998) was observed over 2 L-configuration 
and 1 H-configuration tracks for spatial frequency coverages of $\sim$11-156 
k$\lambda$.  These data were retrieved from the OVRO archive.  In Fall 1998 
and Spring 1999, HCO$^{+}$ 1--0 at 89.1885230 GHz (Pickett et al. 1998) was 
observed over 3 L-configuration tracks, for spatial frequency coverages of 
4-35 k$\lambda$.  These data were also retrieved from the OVRO archive.  In 
Spring 2001 and Fall 2001, N$_{2}$H$^{+}$ 1--0 at 93.1762650 
GHz\footnote{93.1762650 GHz is the frequency of the N$_{2}$H$^{+}$ 101-012 
transition found by Caselli, Myers \& Thaddeus (1995).  The other 6 hyperfine 
components to the 1--0 transition, at frequencies $<$6 MHz from that of 
101-012 were also observed.} (Caselli, Myers \& Thaddeus 1995) was observed 
over 5 L-configuration and 1 H-configuration tracks, for spatial frequency 
coverage of 4-64 k$\lambda$.  These data were obtained specifically for this 
project.  Refer to Paper I for details about observations and data processing. 

\subsection{The Submillimeter Array (SMA) Observations}

L1251B was observed on 25 September 2005 from the summit of Mauna Kea, HI,
with the Submillimeter Array (SMA) in its most compact configuration.  The 
230 GHz SMA receiver was tuned to observe CO 2$-$1, $^{13}$CO 2$-$1, C$^{18}$O 
2$-$1, N$_{2}$D$^{+}$ 3$-$2, SO 5$_{6}$--$4_{5}$, and H$_2$$^{13}$CO 
3$_{12}$--$2_{11}$ in separate correlator windows of various band widths and 
channel spacings.  Table 3 lists the respective frequencies, sidebands of 
observation, correlator windows, band widths, and channel spacings for these 
lines.  Out of the 24 windows in each sideband, the remaining correlator 
windows were used to observe continuum emission over effectively $\sim$2 GHz 
in each sideband.  The continuum data have been presented in Paper I.  Refer 
to Paper I for general information on observations and data processing. 
 
We used the MIRIAD software package for imaging and deconvolution.  For 
the CO and $^{13}$CO data, natural weighting was used without any tapering 
to make the respective data cubes.  The C$^{18}$O and N$_{2}$D$^{+}$ data,
however, were tapered with a circular Gaussian of 4\as\ FWHM during inversion 
to increase the beam size up to $\sim$5\as\ FWHM for improved brightness 
sensitivity.  Furthermore, only 20 velocity channels around the central
velocity were used for the integrated intensity calculation to bring out 
low-brightness features.  Finally, the SO and
H$_{2}^{13}$CO data were tapered with Gaussians of 6\as\ FWHM, also to 
improve brightness sensitivity, but no detections of either line were 
obtained.  To calculate the integrated intensities, all channels were 
used with a 2 $\sigma$ rms clip.

\subsection{Data from Other Observations} 

J. Williams (private communication) provided us with unpublished CS 3$-$2 and 
5$-$4 data that were obtained with the NRAO 12-m telescope\footnote{The Kitt 
Peak 12 Meter telescope was operated by NRAO, a facility of the National 
Science Foundation, operated under cooperative agreement by Associated
Universities, Inc.  It presently operated by the Arizona Radio Observatory 
(ARO), Steward Observatory, University of Arizona, with partial funding from 
the Research Corporation.} on Kitt Peak, AZ in May 1996 (CS 5$-$4) and October 
1997 (CS 3$-$2).  The backend used was the Millimeter AutoCorrelator (MAC) 
with a channel spacing of $\sim$48.8 kHz ($\sim$0.1 \kms\ for CS 2$-$1 and 
$\sim$0.06 \kms\ for CS 5$-$4).  Typical system temperatures were 250-300 K 
(CS 3$-$2) and 450-600 K (CS 5$-$4).  Details of these line observations are 
also summarized in Table 1.
 
\section{RESULTS} 

In this study, we use coordinates in the J2000 epoch, so the ($\alpha$, 
$\delta$) coordinates of IRAS 22376+7455, the IRAS point source associated
with L1251B, are (22$^{h}$38$^{m}$47.16$^{s}$, +75\deg 11\am 28.71\as). 
This position defines the reference center for all figures that use offsets 
in their axes.  For example, IRS1 is located at (0\arcsec, $+$5\arcsec).

\subsection{Observations with Single-Dish Telescopes}

Figure 1 presents the integrated intensity maps of CS 2$-$1, HCN 1$-$0, 
HCO$^{+}$ 1$-$0 and N$_{2}$H$^{+}$ 1$-$0, 
obtained at FCRAO.  In general, these lines trace moderately dense ($\rm n
\leq 3 \times 10^5$ $\rm cm^{-3}$) gas, i.e., they have low excitation 
requirements compared to other lines in this study.  In each panel, line 
emission is shown overlaid onto a map of 850 $\mu$m continuum emission (see 
Paper I) and the locations of the YSOs IRS1, IRS2 and IRS4 are denoted by 
a circle, square and triangle respectively.  Notably, the line maps of Figure 1 
differ from the (low resolution) map of integrated CS 1$-$0 intensity by 
Morata et al. (1997), which instead shows a maximum of emission $\sim$200\as\ 
north of L1251B and no obvious structure around L1251B itself.  These line 
maps are similar to that of NH$_3$ integrated intensity by T$\acute{\rm o}$th 
\& Walmsley (1996), however.  Figure 1 shows that L1251B is coincident with 
line emission maxima in all four tracers, suggesting it is associated with 
moderately dense gas.  Significant line emission is also seen throughout 
the region beyond L1251B, however.  For example, CS 2$-$1 shows a third 
maximum $\sim$130\as\ north of L1251B, but no other tracer in Figure 1 shows 
a maximum towards that general location.  In addition, other maxima are seen 
$\sim$100--160\as\ to the east of L1251B, associated with the ``east core" 
identified from 850 $\mu$m emission in Paper I.  These eastern maxima are 
not coincident with each other, however.  Also, a thin ridge of emission 
extends northward from the east core in all maps.  The brightnesses of the 
maxima are similar for CS 2$-$1, HCN 1$-$0, and H$^{13}$CO$^+$ 1$-$0 (not 
shown) in the east core and L1251B.  The east core, however, is slightly 
brighter in HCO$^{+}$ 1$-$0 but much weaker in N$_{2}$H$^{+}$ 1$-$0 than 
L1251B, although it was not mapped as extensively in the latter line.

Figure 2 presents the integrated intensity maps of CS 3$-$2 and 5$-$4, 
HCO$^{+}$ 3$-$2 and DCO$^{+}$ 3$-$2 obtained with the NRAO 12-m telescope 
or the CSO.  In general, these lines 
trace gas of higher density ($\rm n \geq 1 \times 10^6$ $\rm cm^{-3}$) 
than those shown in Figure 1, i.e., they have relatively high excitation 
requirements.  The maps of integrated intensity in Figure 2 are overlaid 
onto maps of 1.3 mm continuum emission, which traces the column density 
very well.  (As described in Paper I, a shift in the location of the 
continuum emission maximum associated with L1251B to the south is seen 
with increasing wavelength.)  The positions of IRS1, IRS2 and IRS4 are 
shown as in Figure 1, but here ``square-X" symbols denote the locations 
the starless objects from Paper I, SMA-N and SMA-S.  Figure 2 shows that 
L1251B also contains maxima of higher excitation lines, suggesting even 
higher densities than those traced by the lines shown in Figure 1.  
(These maps, however, do not extend as widely as those in Figure 1, so 
the larger-scale distribution of these lines, e.g., toward the east core, 
is not known.)  As in the single-dish continuum data, however, Figure 2 
also shows that the integrated intensity maxima of dense gas tracers are 
located consistently off-center from IRS1.  
For example, the maximum of CS 5$-$4 is located $\sim$15\as\ (greater than a
half beam size) south of IRS1.  
These shifts seen in continuum and molecular emission maps suggest that 
within L1251B the maximum column density, 
and probably the density, resides between IRS1 and IRS2.

Figure 3 shows line profiles at the center position of L1251B of all eight 
lines from Figures 1 and 2, plus H$^{13}$CO$^{+}$ 1$-$0.  The line widths 
of DCO$^+$ 3$-$2, H$^{13}$CO$^+$ 1$-$0 and the isolated component of N$_2$H$^+$
1$-$0, which are obtained by Gaussian 
fits, are $\sim$1 km s$^{-1}$.  The line widths of HCN 1$-$0 and N$_2$H$^+$ 
1$-$0, which are obtained by their hyperfine structure fits, are $\sim$1.4 
and $\sim$1.2 km s$^{-1}$, respectively, 
probably affected by their optical depths. 
The width of CS 5$-$4, however, is 
$\sim$2.6 km s$^{-1}$.  Line wings are seen in the spectra of HCO$^{+}$ 3$-$2, 
CS 5$-$4 and HCN 1$-$0.  All profiles in Figure 3, except for H$^{13}$CO$^+$ 
1$-$0, DCO$^+$ 3$-$2, and the isolated component of N$_2$H$^+$ 1$-$0 are 
asymmetric with blueshifted peak temperatures.  
The peak temperatures of the isolated component of N$_2$H$^+$ 1$-$0 and DCO$^+$ 
3$-$2 lines are located close to the central velocity of 
L1251B, i.e., where the deepest dips are seen in the other lines.  
We determined the central velocities of L1251B ($-3.65$ km s$^{-1}$) and the 
east core ($-4.13$ km s$^{-1}$) by the Gaussian fitting of the isolated 
component of N$_2$H$^+$ 1$-$0 and H$^{13}$CO$^+$ 1$-$0, respectively. 
Figure 4 shows profiles 
of five lines toward the east core, where again asymmetric profiles with 
stronger blue peaks
are seen in HCO$^{+}$ 1$-$0, CS 2$-$1, and HCN 1$-$0 but not H$^{13}$CO 1$-$0. 
Such profiles are not seen everywhere across the region, however, indicating
a complex velocity distribution.  For example, the CS 2$-$1 line profile (not 
shown) has reversed asymmetry toward the northern maximum of CS 2$-$1.  HCN 
1$-$0 (not shown) also has a reversed profile between the east core and 
L1251B.

Figure 5a shows a map of red and blue components of CO 2$-$1 emission toward
L1251B only, from data obtained with the CSO 10.4-m Telescope. 
The velocity ranges for the red and blue components of the outflow have
been determined very conservatively from blue- and red-free spectra 
(Figure 5b).  

The extended 
outflow from L1251B, as seen by others, is plainly noticeable.  The projected 
orientation of the CO outflow is northwest--southeast with blue or red 
components respectively in the northwest or southeast.  The two components 
overlap significantly at L1251B, however, and the origin of the outflow is not
easily identifiable due to the low resolution of the map.

Figure 6 shows HCO$^+$ 3$-$2 spectra from locations across L1251B.  As seen
in Figure 3, the central spectrum shows a very strong self-absorption dip and 
an asymmetric profile with a stronger blue peak.  Figure 6 also reveals that 
the majority of lines off-center to L1251B have asymmetric profiles with 
stronger blue peaks, especially with increasing of 
the asymmetry with a stronger blue peak to the south.  Not all 
the spectra are asymmetric with stronger blue peaks, however.  
To the north, the spectra are 
double-peaked more symmetrically, and in the north, reversed (red) asymmetry 
is seen.  Note that the distribution of HCO$^+$ integrated intensities at 
L1251B (Figure 2c) is elongated through the northwest--southeast direction.
The outflow map (not shown) from the HCO$^+$ 3$-$2 wing components seen in
Figure 3 shows the same trend as seen in the CO 2$-$1 outflow map, i.e., a
northwest--southeast extension and a significant positional overlap between 
the blue and red components.

\subsection{Interferometric Observations}

Figure 7 shows HCO$^+$ 1$-$0, N$_2$H$^+$ 1$-$0, and H$_2$CO 3$_{12}$--2$_{11}$ 
integrated intensities made with data from the OVRO MMA.  Each map in Figure 7 
is overlaid onto the $Spitzer$ IRAC 4.5 $\mu$m band image, and the positions of 
IRS1, etc., are denoted as in Figures 1 and 2.  In these cases, the maps only
show smaller-scale emission features since significant flux on larger scales 
was resolved out by the interferometer.  Figure 7 shows that the distributions 
of compact line emission differ significantly on small scales.  For example, 
Figure 7a shows the HCO$^{+}$ 1$-$0 emission is elongated linearly in the 
northwest--southeast direction from IRS1, similar to the direction of the 
outflow seen on larger scales (see Figure 5).  Figure 7a also shows that the 
N$_{2}$H$^{+}$ 1$-$0 emission is located between IRS1 and IRS2 and is elongated 
in a projected direction perpendicular to that of the HCO$^{+}$ emission.  
In contrast, Figure 7b shows compact H$_{2}$CO 3$_{12}$--2$_{11}$ 
emission in L1251B only at two locations, one centered at IRS1 and another 
2\arcsec\ to the northeast of IRS4.  

Figure 8 shows maps of C$^{18}$O 2$-$1, N$_{2}$D$^{+}$ 3$-$2, CO 2$-$1, and 
$^{13}$CO 2$-$1 integrated intensities from L1251B with data from the SMA.  
Again, the observations were sensitive to only small-scale, compact features.  
The distributions of compact line emission again differ remarkably from each 
other and from those shown in Figure 7.  Figure 8a shows the C$^{18}$O 2$-$1 
emission, along with N$_{2}$H$^{+}$ 1$-$0 emission shown previously in Figure 
7a.  Compact C$^{18}$O emission is found along a filament containing IRS1 and 
IRS2, as well as SMA-N and SMA-S.  Some C$^{18}$O emission is also associated 
with IRS4.  Note, however, that brighter C$^{18}$O emission is slightly offset 
from all objects in L1251B.  N$_{2}$H$^{+}$ emission, however, is distributed 
in a direction perpendicular to the C$^{18}$O filament, but its maxima are 
more coincident with SMA-N and SMA-S (but not completely; see \S 4.4).  In 
contrast, Figure 8b shows compact N$_{2}$D$^{+}$ 3$-$2 emission, which is 
detected in a single location, south of IRS1 and slightly offset to the west 
of SMA-S.  Figures 8c and 8d show the compact CO 2$-$1 and $^{13}$CO 2$-$1 
emission, which both show a lobe-like structure extending southeast of IRS1.  
Bright CO emission is also seen to the northwest of IRS4, with a fainter 
extension coincident with IRS4.  In contrast, bright $^{13}$CO emission is 
coincident with IRS4 but is not seen to the northwest.  Surprisingly, in these 
integrated intensity maps, neither CO nor $^{13}$CO strong emission is seen 
associated with or coincident with IRS2, though its infrared nebulosity 
morphology is suggestive of an outflow cavity.  

\section{ANALYSIS}

According to our observational results, complex dynamical and chemical
processes are coupled in L1251B and the east core.  To study each dynamical 
component, i.e., infall, rotation, and outflow, in detail, we describe in 
this section the analysis of line profiles using line radiative transfer 
calculations and the simulation of observed line profiles (\S 4.1), 
centroid velocity maps and a position-velocity diagram (\S 4.2), and 
channel maps (\S 4.3), respectively.  In addition, we calculate the 
chemical evolution coupled with the evolution of density and luminosity 
to compare the relative distribution of observed various molecular line 
emission with a chemical model (\S 4.4).

\subsection{Infall}

Toward L1251B, optically thick molecular transitions such as HCO$^+$ 1$-$0 
and 3$-$2 show an asymmetrical profile with the brighter blue peak (see 
Figure 3).  In addition, optically thin molecular transitions, such as 
H$^{13}$CO$^{+}$ 1-0 and DCO$^{+}$ 3-2 line are symmetrical, and peak at 
the velocities of the deepest dips of the other lines.  Taken together, 
such profiles can be indicative of infall motions (see Zhou et al. 1993).  
In general, infalling gas will have motions both positive and negative 
along the line of sight, and if an excitation gradient exists also along
line of the sight, redward self-absorption of gas emission can occur.

L1251B was modeled as an inside-out collapsing sphere (cf. Shu 1977) by Young 
et al. (2003), using continuum emission at 450 $\mu$m and 850 $\mu$m to find 
density and temperature profiles of the core.  The model parameters provided
by Young et al., however, do not correspond to the best-fit model (C. Young, 
personal communication).  Instead, the infall radius of the actual best-fit 
collapse model of L1251B is 5000 AU, equivalent to an infall timescale of 
$5\times 10^4$ years, rather than the 3000 AU provided by Young et al.  Other 
parameters of the best-fit model are the same, i.e., an effective sound speed 
of 0.46 km s$^{-1}$ and a total luminosity of 10 \lsun.  The model with a 5000 
AU infall radius provides much better SED fit.  In the current best-fit model, 
the reduced $\chi ^2$ for intensity profiles at 450 $\mu$m and 850 $\mu$m and 
the SED are 17, 27, and 3, respectively (compare with $\chi ^2$ in Table 9 
and 11 of Young et al.) 

Figure 9 shows the density and velocity profiles of the best-fit inside-out 
collapse model. We have used the physical profiles to model HCO$^+$ 3$-$2/1$-$0,
H$^{13}$CO$^+$ 1$-$0 and CS 5$-$4 lines with a Monte-Carlo radiative transfer 
code (Choi et al. 1995). 
These lines trace the densest region, i.e., where the infall velocity is 
significant, among the lines observed with single dish telescopes.  To minimize 
a contribution from rotational motions, line profiles observed only 
at the core center have been modeled.  The kinetic temperature profile (the 
dotted line in Figure 9a) has been calculated with a gas energetics code 
(Young et al. 2004; Doty \& Neufeld 1997) from the dust temperature profile 
obtained from the dust continuum modeling.  Abundance profiles of HCO$^+$ and 
CS were calculated with the chemo-dynamical model developed by Lee et al. 
(2004).  (This chemical calculation and its results will be described in \S 
4.4.)  Although an interferometric observation is necessary to study whether
the high velocity wings of CS 5$-$4 are affected by outflow, Figure 10 shows 
that the density and infall velocity structures from 
the best-fit model of dust continuum observations can reproduce reasonably 
well the observed line profiles detected at the center of the core, especially 
CS 5$-$4.
Note, however, that it was necessary to reduce the abundance profile of 
HCO$^{+}$ by a factor of 2 from that calculated by the chemical model, to
fit the HCO$^{+}$ 3$-$2/1$-$0 line. 
The line profile of H$^{13}$CO$^{+}$ 1$-$0 was also
fitted reasonably well with the HCO$^{+}$ abundance profile and the isotopic
ratio of $^{12}$C/$^{13}$C$=77$ (Wilson and Rood 1994).  
The dotted line in the Figure presents the comparison case without infall.  
In this model, constant turbulent widths of 0.6 \kms\ and 0.25 \kms\ 
(FWHM = 1 and 0.4 \kms, respectively) were assumed at radii respectively less 
than and greater than the infall radius.  Having less turbulent motion in the 
outer stationary envelope than in the inner infalling envelope is necessary 
to fit the width of the self-absorption feature, especially that in HCO$^+$ 
3$-$2.  
The broad line wings of HCO$^+$ 3$-$2/1$-$0 and H$^{13}$CO$^{+}$ 1$-$0, 
particularly pronounced in the blueshifted emission, are probably caused by 
outflowing material, which is not considered in this model.  The higher 
ratio between blue and red peaks of the observed line profiles than the 
modeled ones, which is commonly seen and never understood, 
might be affected by the rotational motion or the non-spherical 
geometry. 

\subsection{Rotation}

Beyond the central position of L1251B, variations of the line profiles and
position-velocity diagrams suggest rotational motions.  For
example, the HCO$^{+}$ 3$-$2 line profiles shown in Figure 6 at positions 
offset from (0,0) show asymmetry reversals and increases that may be caused
by rotational motions (e.g., see Walker et al. 1994; Zhou 1995; Ward-Thompson 
\& Buckley 2001).  A simulation of the HCO$^+$ 3$-$2 spectra with a 2-D Monte 
Carlo molecular line radiative transfer code (Hogerheijde \& van der Tak 2000) 
for the collapse from a rotating dense cloud (Terebey, Shu \& Cassen 1984) 
indeed produces the distribution of blue- and red-asymmetric profiles observed 
in L1251B (Lee et al., in preparation).  Note, however, that the wings of
line profiles along the projected outflow direction may be still affected by 
outflow motions (see \S 4.3 below).

Previous studies (Goodman et al. 1993; Sato et al. 1994; T$\acute{\rm o}$th
\& Walmsley 1996; Anglada et al. 1997; Morata et al. 1997) determined that
the overall velocity gradient in the L1251B neighborhood, i.e., that covered
by the extended 850 \micron\ map in Figure 1, was $\sim$1-2 km s$^{-1}$ 
pc$^{-1}$ in the northeast--southwest direction.  
Observations by Caselli et al. (2002),
however, revealed a more complex velocity distribution in the neighborhood,
with gradients of opposite direction.  Namely, L1251B itself is associated
with a southwest--northeast gradient, while the east core is associated with a
northeast--southwest gradient.  The average direction reported in their paper,
however, is also northeast--southwest, suggesting that the previous studies,
with poorer resolutions, averaged out the velocity gradient associated with
L1251B, whose velocity gradient is only about half of that in the east core.
To determine the velocity gradient associated with L1251B and the east core
at higher resolution,
we fitted the hyperfine structures of HCN 1$-$0 and N$_2$H$^+$ 1$-$0 and the
Gaussian profile of H$^{13}$CO$^+$ 1$-$0, and found the same result as shown
in Figure 6 of Caselli et al.  For illustration, Figure 11 shows the centroid
velocity distribution obtained from HCN 1$-$0, which shows different velocity
gradients in L1251B and the east core.  The derived velocity gradient from
HCN 1$-$0 in L1251B within 80\arcsec\ is $\sim 3/{\rm cos}(i)$ km s$^{-1}$ 
pc$^{-1}$ in the southwest--northeast direction but the velocity gradient in 
the east core is $\sim 6/\rm{cos}(i)$ km s$^{-1}$ pc$^{-1}$ in the opposite 
direction, i.e., consistent with the direction predicted from previous studies. 
Here, $i$ is the inclination of the rotational axis from the plane of sky. 
 
We have analyzed the centroid velocity shift in L1251B with the
N$_{2}$H$^{+}$ 1$-$0 interferometer data (see Figures 7a and 8a).
Figure 12a shows a 2-D distribution of the mean centroid velocity with  
the integrated intensity contours of the isolated component of N$_{2}$H$^{+}$ 
1$-$0 , which were calculated with the AIPS task, MOMNT. 
A clear shift of the centroid velocity is seen from the southwest to the 
northeast, along a direction perpendicular to the outflow direction. 
This southwest--northeast direction is {\it opposite} to directions of
velocity gradient seen in previous studies with lower resolution
but consistent with the result of Caselli et al. (2002) and that seen in 
HCN 1$-$0 (Figure 11) toward L1251B.
A position-velocity diagram taken along a cut centered at (9\as, $-2$\as) and 
along P.A. = 33\degree\ (similar to the direction of maximum elongation of 
the N$_{2}$H$^{+}$ emission) is seen in Figure 12b.
The rotational motion suggested by this gradient
is very fast, i.e., $30/cos(i)$ km s$^{-1}$ pc$^{-1}$, $\Omega \sim
10^{-12}/cos(i)$ s$^{-1}$ within 30\arcsec.  
This gradient is $\sim$1 order of magnitude larger than what is seen 
on larger scales, such as in the HCN map (Fig. 11), suggesting an 
increased importance of rotation on small scales. 

\subsection{Outflow}

In this section, we examine at high angular resolution the outflowing gas 
from the young stellar objects in L1251B, and provide an initial interpretation 
of the data.  Although the evidence for outflows in this region is strong, the 
spatial distribution of this gas is very irregular, making it difficult to 
associate the observed outflow features with specific objects.

Figure 13 shows the CO 2$-$1 integrated intensity across L1251B as observed
from the SMA, divided into red and blue components (emission at line center
is excluded).  For reference, the CO components are overlaid onto the 1.3 mm 
continuum emission across L1251B, also observed from the SMA.  Figures 14 
and 15 show channel maps of CO 2$-$1 respectively for red- and blueshifted 
emission.  For reference, the CO emission in each channel is overlaid here
onto the N$_{2}$H$^{+}$ 1-0 integrated intensity as observed from the OVRO 
millimeter array.  (Note that the velocity range definitions of red- and 
blueshifted emission are slightly different in these Figures than those used 
for Figures 5 and 13 because of the difference in their velocity resolutions.)
In Figures 14 and 15, we also include more channel maps at -1.2 \kms, and 
-5.4 and -6.5 \kms, in the red- and blueshifted emission, respectively, to 
resolve different outflow components. 
For an alternative view of the outflows associated with 
L1251B, Figure 16 shows channel maps of HCO$^{+}$ 1-0 observed from OVRO.
In the channel maps, especially for blueshifted emission (Figure 15 and 16),  
velocities closer to the central velocity than defined conservatively in 
Figure 5 and 13 are covered. 
While the HCO$^{+}$ 1-0 emission around the central velocity peaks at SMA-S, 
which is the densest part in L1251B, at -5.2 and -5.6 \kms, it has peaks at 
IRS1 and southeast of IRS4 with a weaker tail toward the northwest of IRS4.
Therefore, HCO$^{+}$ 1-0 mainly traces the outflowing material even at the 
velocity up to $\sim -5$ \kms.
In addition, the comparison between the model and observation of HCO$^{+}$ 1-0
in Figure 5b suggests that the emission at -5 \kms\ is affected only by 
the outflow.   

As can be seen from Figure 13, significant outflow appears to be associated 
with IRS1.  This emission, however, may also include a component from SMA-S.  
Bright emission is seen extending southeast of IRS1, with blueshifted emission 
to the east-southeast and redshifted emission to the south-southeast and some 
positional overlap of these components to the southeast.  The overlap of the 
components could indicate a single outflow where the inclination of the axis
from the plane of sky is not large and the blue and red components respectively
trace the front and back components of an outflow cone.  Note, however, that 
in the redshifted -1.2 \kms\ channel of Figure 14, the feature divides into
two components, one associated to the west and southwest of IRS1 and another
to the southwest of SMA-S.  Note also the pair of distinct components seen in 
Figure 15, associated with IRS1 in one blueshifted pair of channels at -5.4 
\kms\ and -6.5 \kms\ and associated with SMA-S in another blueshifted pair of 
channels at -9.6 \kms\ and -11 \kms.  In the latter case, the emission may 
have been made compact and elongated in the northwest-southeast direction 
due to higher densities to its immediate northeast, as evident from the
N$_{2}$H$^{+}$ emission and the higher-excitation line emission (see Figure 
2). 

For IRS2, a single redshifted feature is seen extending to its southeast in 
Figure 13, and this feature is also seen in the CO channel maps from 2 \kms\ 
to -1.2 \kms\ of Figure 14.  No corresponding blueshifted feature is obviously 
associated with IRS2, however.  It is possible that the blueshifted feature 
near IRS1 seen at -5.4 \kms\ and -6.5 \kms\ is associated with IRS2 instead.
For example, CO may be severely depleted in the dense regions between IRS1 
and IRS2, and the blueshifted outflow from IRS2 may be only visible in CO 
near IRS1 because of localized high CO abundance due to the evaporation of 
CO from nearby dust grains.  Note, however, that this blueshifted emission 
lies in a similar northwest-southeast direction as redshifted emission from 
IRS1.  Alternatively, the extended redshifted feature southeast of IRS2 may 
be related to IRS1 itself.

Northwest of IRS4, Figure 13 shows red- and blueshifted features with very
significant positional coincidence.
Again, these features could be due to a single outflow of 
low inclination.  The channel maps of Figure 14 and 15 suggest a more complex 
interpretation, however, with weak red- and blueshifted emission seen both 
northwest and east of IRS4 at -1.2 \kms, -5.4 \kms\ and -6.5 \kms.  Figure
16 shows blueshifted emission from HCO$^{+}$ 1-0 also located to the east 
and the northwest of IRS4.  Instead, there may be two outflows present, one 
centered at IRS4 and another to the northwest.  This latter may originate 
from IRS1 and could be the counterpart to the redshifted emission seen 
southeast of IRS2.  Note that not much material is seen between IRS4 and 
IRS1, in the continuum emission maps of Figures 2 and 9 or the CO 2$-$1 
channel maps of Figure 13.  Compact HCO$^{+}$ 1$-$0 emission {\it is}\/ 
seen between IRS4 and IRS1, however, in Figure 16 from -6 \kms\ to -5 \kms.  
The strongest HCO$^{+}$ emission is seen southeast of IRS4, and this may
constitute a blueshifted outflow component from IRS4. 

A summary of the possible outflows in the L1251B region is shown in Figure 
17, with outflow components schematically plotted over the 1.3 mm continuum
emission observed with the SMA.  In this picture, IRS1, IRS2, IRS4 and even
SMA-S all have outflows, some of which interact with each other.  If SMA-S
indeed has an outflow, it is not a starless condensation after all.  Instead,
it may harbor a very low luminosity protostellar object that was undetected
by Spitzer.  The existence of such objects has been surmised by others; e.g., 
Rebull et al. (2007) suggested that several objects unseen at $\lambda$ 
$<$ 70 $\mu$m could be driving numerous outflows observed in HH 211.  Future 
observations of L1251B, at even higher resolutions or using various lines that 
clearly trace gas shocked by outflow motions (e.g., mid-infrared lines of 
H$_{2}$ or millimeter lines of SiO), will be very helpful in disentangling 
the association of the outflows with specific sources and their effects on 
the evolution of the L1251B core.

\subsection{Chemistry}

Within L1251B, our high-resolution maps provide continued evidence for the 
expected chemical behavior of dense gas not associated with protostars.  For
example, the HCO$^{+}$ features shown in Figure 7a likely traces outflowing 
dense gas, since HCO$^{+}$ can be made abundant through shocks (see Rawlings
et al. 2004).  As with HCO$^{+}$, the CO, $^{13}$CO, and diffuse C$^{18}$O 
emission seen in Figures 8a, c and d may also be due to extensive depletion 
in the extended envelope and enhancements along the outflow axis due to 
liberation from grains.  Emission from these lines indeed share some common 
features, but note that no compact CO, $^{13}$CO, or C$^{18}$O emission is 
seen between IRS1 and IRS4.  A faint ridge of HCO$^{+}$ 1$-$0 emission is 
seen at that location in the level of 2 $\sigma$, however.   

In addition, the N$_{2}$H$^{+}$ feature shown in Figure 7a likely traces 
non-outflowing gas, since N$_{2}$H$^{+}$ can remain abundant in cold dense 
material and can be depleted significantly in outflows.  N$_2$H$^+$ is a 
chemical daughter of N$_2$, which forms slowly.  The interactions between 
N$_2$H$^+$ and grains replenish N$_2$ in the gas phase because N$_2$H$^+$ 
recombines with electrons on the grain surfaces.  The principal destroyers 
of N$_2$H$^+$ in the gas phase are CO and electrons.  CO can be significantly 
depleted in starless cores (Bergin \& Langer 1997) because the only source 
of heating is the interstellar radiation field and the inner temperatures of
such cores can be correspondingly very low (e.g., $\sim$7 K; see Evans et al. 
2001).  Also, the electron abundance can be very low ($\sim 10^{-9}$, Williams 
et al. 1998) in dense gas that is highly extincted, like that in starless 
cores.  The N$_2$H$^+$ emission seen in Figure 7a and 8a is consistent with 
the 1.3 mm dust continuum emission mapped by the IRAM 30-m Telescope (see 
Paper I), suggesting it is also tracing cold, dense gas.  

N$_2$H$^+$ (and N$_2$D$^+$) will also deplete at densities greater than 
$10^6$ cm$^{-3}$ (Di Francesco, Andr\'e \& Myers 2004; Pagani et al. 2005).  
Note that the two N$_2$H$^+$ intensity maxima in Figure 7a and 8a are offset
slightly to the east from SMA-N and SMA-S, which trace the densest parts of 
L1251B.  (The C$^{18}$O emission observed with the SMA does not have maxima 
at SMA-N and SMA-S either.)   In addition, an N$_{2}$D$^{+}$ emission maximum 
is found between IRS1 and SMA-S.  Finally, no such emission is associated 
with SMA-N.  The offset of the N$_2$H$^+$ maxima and deficiencies of 
N$_{2}$D$^{+}$ from these objects are possibly caused by the depletion from 
the gas phase at densities greater than $10^6$ cm$^{-3}$.  The N$_2$D$^+$ 
emission may be identifying a less dense location that has a high enough 
temperature to populate significantly N$_2$D$^+$ at J$=$3.  For example, the 
J$=$3 level of N$_2$D$^+$ has its maximum population in thermal equilibrium 
at $\sim$20 K, and this may be caused by heating by IRS1.  (Note, however, 
that the temperature there cannot be greater than 30 K, because otherwise 
CO would evaporate off dust grains and reduce the N$_2$D$^+$ abundance.)

Our high-resolution observations of L1251B show further examples of the
chemical behavior expected in the presence of nearby protostellar heating.  
Once a protostar forms, the surrounding material is heated up to the CO 
evaporation temperature ($\sim$25 K in the case of bare SiO$_2$ dust grains).  
The desorbed CO destroys N$_2$H$^+$, producing an N$_2$H$^+$ emission 
``hole" around the central heating source, that is potentially observable 
with interferometers.  Such holes may explain the lack of N$_{2}$H$^{+}$ 
emission coincident with IRS1 and IRS2, as seen in Figures 7a and 8a.  In 
contrast, the abundances of other molecules such as CS, H$_2$CO, and HCO$^+$ 
can be enhanced due to protostellar heating resulting in the desorption of 
CO or the desorption of themselves off of dust grain mantles (Lee, Bergin 
\& Evans 2004) as seen in Figure 7b.  In L1251B, IRS1 is luminous enough 
to evaporate CO in its inner envelope.  Similar N$_2$H$^+$ emission holes 
have been seen toward NGC 1333 IRAS 4A (Di Francesco et al.  2001) and L483 
(J{\o}rgensen 2004).

Figure 18 shows the results of a chemical evolution model made specifically 
for L1251B to quantify the observed line emission distributions.  For this 
model, we used the chemo-dynamical model developed by Lee et al. (2004).
We updated the chemical network to include more recent results on the 
N$_2$H$^+$ chemistry, however, including a new binding energy of N$_2$, 
which is the same as that of CO (\"Oberg et al. 2005), and new rates of 
dissociative recombination of N$_2$H$^+$ with electrons (Geppert et al. 
2004).  We also increased the initial abundance of sulfur by a factor of 
3 to fit the CS 5$-$4 line profile in Figure 10a.  The initial abundances 
of other species are the same as those in Table 3 of Lee et al. (2004).  
For a dynamical model, we adopted the best-fit Shu inside-out collapse 
model to the dust continuum emission as described in \S 4.1.  Although 
L1251B has several sources, only one internal luminosity source at the 
center of a spherically symmetric envelope was assumed in the model.  
Since IRS1 is the dominant luminosity source by a factor of $\sim$10, 
however (see Paper I), this model is a reasonable first approximation, 
especially for interpreting the single dish observations.  For this model, 
the infall rate from the disk to central protostar was tuned appropriately 
to match the luminosity calculated by observations at the given timescale 
(Young \& Evans 2005).  In addition, the interstellar radiation field was 
assumed to be attenuated to $G_0=0.3$ for consistency with the dust 
continuum modeling.  An outflow was not included.  

We do not compare this chemical model with interferometric observations 
quantitatively since the density profile assumed in the 1-D model is not 
appropriate for the high resolution observations that reveal multiple sources.
The chemical distribution close to the central source, however, is most 
sensitive to the temperature environment, so we can look for the effects
of temperature increases around IRS1 on the chemical abundances.  As seen 
in Figure 18, the model predicts a CO evaporation radius (which accordingly 
is also the N$_{2}$H$^{+}$ depletion radius) of $\sim$0.007 pc ($\sim$4\as), 
i.e., similar to the observed radius of the N$_2$H$^+$ hole at a 2 $\sigma$ 
level of integrated intensity (see Figure 7a).  Furthermore, the model 
predicts an H$_2$CO abundance peak at a radius of $\sim$0.003 pc ($\sim$2\as), 
similar to the radius of the H$_2$CO emission at the 2 $\sigma$ level of 
the integrated intensity (Figure 7b).  A second H$_2$CO abundance peak seen 
in Figure 18 caused by CO evaporation does not affect significantly the 
H$_2$CO emission distribution likely because of the lower density and 
temperature of material at $\sim$0.006 pc relative to material at $\sim$0.003 
pc associated with the inner abundance peak.  Therefore, this simple model
is in good agreement with our observations close to the central source, 
where infall is kinematically dominant, and chemistry mainly depends on the 
evaporation of molecules from grain surfaces by heating.   

Near IRS4, the associated H$_{2}$CO emission may not be due to an increased 
abundance due to localized dust grain heating.  That source may not be 
luminous enough to heat grains above the H$_{2}$CO evaporation temperature 
at a projected distance of 600 AU.  The coincidence of the maxima of the 
integrated intensity of H$_{2}$CO 3$_{12}$--2$_{11}$ (see Figure 7b) to 
one of that of HCO$^{+}$ 1-0 (see Figure 7a) suggests a possible origin 
from shocked outflow material.  Note that the H$_{2}$CO emission close to 
IRS4 lies at redshifted velocities of $-1.5$ \kms\ to $-3$ \kms, and lies 
spatially at a full synthesized beam width (2\arcsec) northeast of IRS4.

\section{DISCUSSION} 

\subsection{L1251B}

L1251B harbors a small group of starless and protostellar objects.  Three 
of the latter are classified as Class 0/I candidates, which are associated 
with dense envelopes.  The observations of molecular lines around L1251B 
have revealed active processes, either dynamical (infall, rotation and 
outflow) or chemical (depletion and enhancement). 

The 1.3 mm dust continuum observations of L1251B with the SMA revealed two 
condensations (SMA-N and SMA-S) unassociated with any detected near-infrared 
source.  (Note that SMA-S may contain a newly formed protostellar object if 
an outflow originates from it; see Figs. 14 and 15, and \S 4.3).  
Table 4 shows masses along the line of sight towards IRS1, IRS2, SMA-N and
SMA-S, determined using the peak intensities of the 1.3 mm image and assuming
a dust temperature of 20 K.  We determined masses along the lines of sight
rather than total masses from the respective fluxes because the extents of
each were difficult to define with certainty in the crowded field, especially
for SMA-N and SMA-S.  Furthermore, any differences in mass along the line of
sight could be more easily discerned with a single sampling.  Table 4 shows
that the peak intensities are quite similar (i.e., within 50\%), suggesting
quite similar masses along the line of sight if the temperatures are common.
If we assume for SMA-N and SMA-S a lower temperature of 10 K, which is more
commonly adopted for starless cores, their masses along the line of sight are
greater than those for 20 K (and IRS1 and IRS2) by a factor of $\sim$3.

Given their designations as protostars, the fluxes 
of IRS1 and IRS2 likely include emission from their respective disk components. 
In particular, IRS2 is not resolved at the $\sim$4\as\ resolution of the data, 
which is equivalent to a $\sim$1200 AU linear distance at the 300 pc distance 
of L1251B.  Since these sources are also highly embedded, however, the 1.3 mm 
fluxes also likely contain emission from their inner envelopes.  In contrast, 
the fluxes of SMA-N and SMA-S likely include emission only from envelopes since 
these sources are not seen if lower spatial frequences (i.e., $<$11 k$\lambda$) 
are excluded, as discussed in Paper I.  To distinguish the disks of IRS1 and 
IRS2 (as well as a possible disk of SMA-S) associated with outflows (\S 4.3), 
observations with higher resolution are needed.  Note that 1.3 mm continuum
emission is not detected towards IRS4, suggesting that it is not associated 
with any disk or inner envelope that could be detected at our sensitivity and 
resolution.

All objects in L1251B possibly form by fragmentation during gravitational 
collapse (Boss 1997; Machida et al. 2005).  Turbulent fragmentation seems 
unlikely since the lines observed toward L1251B are not so broad as to be 
considered dominated by turbulence.  For example, tracers of very dense gas 
such as DCO$^+$ 3$-$2, N$_2$H$^+$ 1$-$0, and H$^{13}$CO$^+$ 1$-$0 have line 
widths of $\sim 1$ km s$^{-1}$, consistent with the infall velocity at $n 
\sim 10^6$ cm$^{-3}$ in the best fit inside-out collapse model of L1251B.  
In addition, CS 5$-$4, which traces much higher densities ($n \geq 5\times 
10^6$ cm$^{-3}$), shows the broadest line width among all the high density 
tracers we observed toward L1251B, i.e., $\sim 2.6$ km s$^{-1}$.   Indeed, 
our inside-out collapse model shows an infall velocity at $5\times 10^6$ 
cm$^{-3}$ of $\sim 2$ km s$^{-1}$.  Also, the infall velocity profile this 
model fits the observed CS 5$-$4 and HCO$^+$ 3$-$2 reasonably well, with 
constant turbulent widths of 0.6 \kms\ and 0.25 km s$^{-1}$ at radii less 
than and greater than the infall radius respectively (see Figure 10a).  The 
velocity dispersions implied from such widths are not much greater than the 
thermal velocity dispersion before collapse.  (The dotted line in Figure 10 
presents the case without the infall velocity structure.)  Therefore, the 
dynamics of L1251B seem dominated by gravitational collapse, combined with 
rotation and outflows, rather than turbulent motions.

\subsection{The East Core}

The east core seen in Figure 1 at 850 \micron\ and in various molecular lines 
has a radius of about 0.1 pc.  The 850 $\mu$m emission shows several sub-cores 
inside a larger structure traced by molecular emission.  This difference may 
be related to a bias towards imaging smaller-scale emission in the continuum 
data and related to high optical depths in the line data.  For the continuum 
data, the SCUBA observations were made by chopping onto the sky, effectively 
filtering out emission on angular scales greater than the $\sim$120$\arcsec$ 
chop throw (see Di Francesco et al. 2007), i.e., those traced by the molecular 
line data. 
In addition, the 850 \micron\ continuum emission is very likely
optically thin, tracing quite dense material within a molecular core.  For the 
line data, the molecular emission (see Figure 1) may have high optical depths 
and the sub-cores may not be visible, despite the similarities between the 
densities these particular lines trace and the mean density of the sub-cores, 
i.e., $\sim$10$^{5}$ cm$^{-3}$.  Differences in the positions of maximum 
integrated intensities between these lines may also be due to their relative 
(but high) optical depths. 
(To determine if the $\sim$3$\times$ higher resolution of the 850 $\mu$m map
relative to the molecular line maps in Figure 1 contributed to the different
appearance of the continuum and line maxima, we smoothed the 850 $\mu$m map
to a resolution similar to that of the line maps (60\as).  The smoothed 850
$\mu$m map, however, still shows an emission hole near the maxima of the line
maps.)

The low optical depth of the 850 \micron\ emission suggests that the much
stronger continuum peak toward L1251B ($\sim$1 Jy beam$^{-1}$) relative to 
the east core ($\sim$0.2 Jy beam$^{-1}$) is indicative of a larger column
density toward L1251B even if L1251B has been heated by internal sources.
(Note that the 850 \micron\ emission peaks close to SMA-S, not toward IRS1,
as seen in Figure 6 of Paper I) 
The 850 $\mu$m fluxes of L1251B and the whole east core are similar, 
suggesting that
their masses are also similar if their temperatures and dust opacities are 
themselves similar, i.e., $\sim 4$ Jy and $\sim 2$ \msun\ if T = 20 K and
$\kappa_{1.3 mm}$ = 0.02 cm$^{2}$ g$^{-1}$ (Kr\"ugel \& Siebenmorgen 1994).
Since L1251B contains several YSOs but the east core does not, it is likely 
that L1251B is warmer than the east core, however.  Continuum observations 
at multiple wavelengths and modeling are needed to determine better relative 
values of dust temperature.

Table 5 lists the integrated intensities of molecular lines observed toward 
the centers of L1251B and the east core.  To probe for any relative chemical 
evolution in the two cores, we compare the relative integrated intensities of 
the 1$-$0 transitions of H$^{13}$CO$^+$ 1$-$0 and N$_{2}$H$^{+}$, molecules 
found to be abundant respectively at earlier and later times in models of 
chemical evolution (e.g., Lee et al. 2003 and references therein).  For 
H$^{13}$CO$^+$ 1$-$0, the integrated intensities of L1251B and the east core 
are similar (i.e., $\sim$1.3 K \kms\ and $\sim$1.1 K km s$^{-1}$ respectively).
For N$_{2}$H$^{+}$, however, the integrated intensity towards L1251B is larger 
than that towards the east core by a factor of two (i.e.,  $\sim$4 K km 
s$^{-1}$ vs $\sim$2 K km s$^{-1}$ respectively).  The relative integrated 
intensity ratio between these lines, i.e., $\sim$6 for L1251B and $\sim$3 for
the east core, may be due to the relative chemical evolution, with L1251B 
being more chemically evolved.  (Note that the difference in N$_{2}$H$^{+}$ 
abundance might be greater than the difference in the N$_{2}$H$^{+}$ 1$-$0 
intensity due to the higher optical depth in L1251B.)  Such differences in 
chemical evolution may result from differences in density since the chemical 
timescale shortens at a higher density.  In addition, H$^{13}$CO$^{+}$ and its 
isotopologue HCO$^{+}$ deplete significantly more than N$_{2}$H$^{+}$ in cold, 
dense environments such as in L1251B (except in the immediate vicinities of 
protostellar objects).  Therefore, the higher ratio between the N$_{2}$H$^{+}$
and H$^{13}$CO$^{+}$ intensities in L1251B than 
in the east core (Figure 1) may be caused by a denser environment in L1251B.  

Various lines toward the east core suggest infall motions.  Therefore, the 
east core may be in earlier states of dynamical and chemical evolution than 
those of L1251B, and may harbor star formation in the future. 

\section{SUMMARY}

L1251E, the densest C$^{18}$O core of L1251 (Sato et al. 1994), has been 
observed at various molecular line transitions with single-dish telescopes, 
and L1251B, the protostellar group in the C$^{18}$O core, has been observed 
with millimeter interferometers.  We have compared these molecular line data 
to continuum data presented in Paper I to study this region of L1251 more 
comprehensively.  Our results include:

1. L1251E contains at least two cores, one
coincident with L1251B and an east core detected from 850 \micron\ continuum 
emission and various emission lines.  The integrated intensities of these 
lines show primary maxima that are coincident with L1251B and other, secondary 
maxima associated with the east core which are non-coincident.  No embedded 
sources have been detected in the $Spitzer$ IRAC or MIPS bands toward the 
east core (Paper I), suggesting it is starless.  Asymmetric profiles with 
stronger blue peaks have been detected toward the eastern core, however, that 
are suggestive of infall motions.  The east core appears less dense and less 
chemically evolved than L1251B.

2. The large-scale outflow associated with L1251B has been mapped in CO 
2$-$1 with the CSO.  Red and blue components of the outflow overlap around 
the brightest protostar, IRS1, and the near-infrared bipolar nebula source, 
IRS2.  The CO 2$-$1 and $^{13}$CO 2$-$1 maps observed with the SMA 
resolve the outflow into a few components.

3. HCO$^+$ 3$-$2 line emission observed with the CSO across L1251B shows 
widespread line asymmetry that suggests infall motions towards the central 
group, with evidence for rotational or outflow motions from modifications 
of the line profiles at many off-center locations.  

4. On a large scale, L1251B has a velocity gradient that is half 
the magnitude of that in the starless east core and is opposite in direction.  
Previous low-resolution observations averaged the velocity gradients of both, 
causing rotational motions in the region to be misinterpreted.

5. HCO$^+$ 1$-$0 line emission observed with the OVRO MMA is centered at IRS1 
and distributed linearly along the northwest--southeast direction of the 
larger-scale outflow.  In contrast, the N$_2$H$^+$ 1$-$0 line emission also 
observed with the OVRO MMA is distributed between IRS1 and IRS2 (like the 
position of maximum brightness of the IRAM 1.3 mm continuum emission) and 
extended
perpendicular to the outflow, i.e,. along the southwest--northeast direction. 
A position-velocity diagram of N$_2$H$^+$ 1$-$0 indicates very fast rotation 
in this flattened envelope, i.e., $\Omega \sim 10^{-12}/cos(i)$ s$^{-1}$, 
where $i$ is the inclination of the rotational axis from the plane of sky.  

6. The maxima of N$_2$H$^+$ 1$-$0 and N$_2$D$^+$ 3$-$2 integrated intensity
observed with the OVRO MMA and SMA are offset from locations of maximum 
continuum brightness from the two starless objects, which were newly detected 
at 1.3 mm with the SMA (Paper I).  This offset may be due 
to reduction of N$_2$H$^+$ and N$_2$D$^+$ abundances at very high densities.

7. OVRO MMA and SMA observations reveal a lack of N$_2$H$^+$ emission at
IRS1 (i.e., a ``hole") but C$^{18}$O and the H$_2$CO emission is coincident 
with IRS1.  These results are consistent with a chemo-dynamical evolution 
model where a central protostar heats up surrounding material.  In this model, 
CO and H$_{2}$CO are abundant close to the protostar where they are released 
into the gas phase from grain mantles, but N$_{2}$H$^{+}$ has had its 
abundance sharply reduced since it is depleted by reactions with gas-phase
CO.

\acknowledgments
Support for this study was provided by NASA through Hubble Fellowship grant 
HST-HF-01187 awarded by the Space Telescope Science Institute, which is 
operated by the Association of Universities for Research in Astronomy, Inc.,
for NASA, under contract NAS 5-26555.  This work was also supported by NSF 
grant AST-0307350 to the University of Texas at Austin and by the State of 
Texas.  We are very grateful to an anonymous referee for valuable comments.
We thank Jonathan Williams for allowing us to use his unpublished 
molecular line data, and we also thank Chad Young for his help for the dust 
continuum modeling of L1251B.  
We are also very grateful to Anneila Sargent for 
providing us the opportunity to observe L1251B with the OVRO MMA.

%%%%%%%%%%%%%%%%%%%%%% References %%%%%%%%%%%%%%%%%%%%%%%%%%%%%%%%%%

\clearpage

%%%%%%%%%%%%%%%%%% Table 1 %%%%%%%%%%%%%%%%%%%%%%%%%
\begin{deluxetable}{lcccccc}
\tablecolumns{7}
\footnotesize
\tablecaption{\bf Molecular line observations with single-dish telescopes  
\label{tab1}}
\tablewidth{0pt}
\tablehead{
\colhead{Line}                &
\colhead{$\nu$}    &
\colhead{$\delta v$}              &
\colhead{$\theta_{mb}$}              &
\colhead{$\eta_{\rm mb}$\tablenotemark{a}}              &
\colhead{Observing Dates}              &
\colhead{Observatory}              \\
\colhead{}                      &
\colhead{(GHz)}        &
\colhead{(km s$^{-1}$)}                      &
\colhead{(arcsec)}                      &
\colhead{}                  &
\colhead{}                  &
\colhead{}
}
\startdata
H$^{13}$CO$^+$ 1$-$0 & 86.754285 & 0.14 & 62 & 0.5 & April 1997, May 2000 & 
FCRAO \\
HCN $1-0$ & 88.631847 & 0.10 & 61 & 0.5 & February 2005 & FCRAO \\
HCO$^+$ 1$-$0 & 89.188523 & 0.13 & 60 & 0.5 & April 1997, May 2000 & FCRAO \\
N$_2$H$^+$ 1$-$0 & 93.176265 & 0.13 & 58 & 0.5 & December 1996 & FCRAO \\
CS $2-1$ & 97.980968 & 0.10 & 55 & 0.5 & February 2005 & FCRAO \\
DCO$^+$ $3-2$ & 216.112605 & 0.13 & 34.7 & 0.74 & July 2003 & CSO\\
CO $2-1$ & 230.53800 & 0.12 & 32.5 & 0.64 & July 2003 & CSO\\
HCO$^+$ 3$-$2 & 267.557526 & 0.12 & 22.5 & 0.67 & June 1997, July 1998 & CSO\\
CS $3-2$ & 146.969048 & 0.10 & 43 & 0.8\tablenotemark{b} & October 1997 & NRAO\\
CS $5-4$ & 244.935606 & 0.12 & 26 & 0.5\tablenotemark{b} & May 1996 & NRAO \\
\enddata
\tablenotetext{a}{
This main beam efficiency is used to convert the antenna temperature ($T_a^*$), 
only corrected for atmospheric attenuation, radiative loss and reward 
scattering and spillover, to the radiation temperature. 
}
\tablenotetext{b}{
These numbers represent the corrected main beam efficiencies ($\eta_{\rm m}^*$) 
to convert the antenna temperature ($T_R^*$), corrected for forward scattering 
and spillover as well as atmospheric attenuation, radiative loss and reward 
scattering and spillover, to the radiation temperature.
}
\end{deluxetable}

\clearpage

%%%%%%%%%%%%%%%%%% Table 2 %%%%%%%%%%%%%%%%%%%%%%%%%
\begin{deluxetable}{cccccccc}
\rotate
\tablecolumns{8}
\tablewidth{0pc}
\tablecaption{OVRO Observational Summary}
\tablehead{
\colhead{} & \colhead{} & \colhead{} & \colhead{Channel} & \colhead{Gaussian} &
\colhead{Synthesized} & \colhead{Synthesized} \\
\colhead{} & \colhead{} & \colhead{Bandwidth} & \colhead{Spacing} &
\colhead{Taper FWHM} & \colhead{Beam FWHM} & \colhead{Beam P.A.} &
\colhead{1 $\sigma$ rms\tablenotemark{a}} \\
\colhead{Tracer} & \colhead{Transition} & \colhead{(MHz)} &
\colhead{(km s$^{-1}$)} & \colhead{(\as\ $\times$ \as)} &
\colhead{(\as\ $\times$ \as)} & \colhead{(\deg)} & \colhead{(Jy beam$^{-1}$)}}
\startdata
H$_{2}$CO  &      3$_{12}$--2$_{11}$ & 7.00  & 0.17     & 1.60 $\times$ 1.60  &
2.5 $\times$ 2.3   &  304  &  0.1    \\
HCO$^{+}$  &      1--0               & 8.00  & 0.42     & 5.5 $\times$ 5.5    &
 7.5 $\times$ 6.9  &  313  &  0.09   \\
N$_{2}$H$^{+}$ &  1--0               & 7.50  & 0.20     & 4.0 $\times$ 4.0    &
 6.5 $\times$ 5.2  &  334  &  0.07   \\
\enddata
\tablenotetext{a}{1 $\sigma$ rms computed from signal-free regions of the
deconvolved maps.}
\end{deluxetable}

%%%%%%%%%%%%%%%%%% Table 3 %%%%%%%%%%%%%%%%%%%%%%%%%
\begin{deluxetable}{lccccccc}
\tablecolumns{8}
\footnotesize
\tablecaption{\bf SMA Line Observations
\label{tab1}}
\tablewidth{0pt}
\tablehead{
\colhead{Line}  &
\colhead{Freq.\tablenotemark{a}}  &
\colhead{S-band}  &
\colhead{Window}  &
\colhead{Chans}  &
\colhead{$\rm \Delta V$}  &
\colhead{Beam FWHM}  &
\colhead{P.A.}\\
\colhead{}  &
\colhead{(GHz)}  &
\colhead{}  &
\colhead{}  &
\colhead{}  &
\colhead{(km s$^{-1}$)}  &
\colhead{(\arcsec x \arcsec)}  &
\colhead{(degree)}
}
\startdata
CO 2--1            & 230.5380000&USB  & 14& 128 & 1.06 & 4.1 $\times$ 3.1& -26.2\\
$^{13}$CO 2--1     & 220.3986765&LSB  & 13& 512 & 0.26 & 4.3 $\times$ 3.4& -23.5\\
C$^{18}$O 2--1     & 219.5603568&LSB  & 23& 512 & 0.26 & 5.3 $\times$ 5.0
\tablenotemark{b}& 0.50\\
N$_{2}$D$^{+}$ 3--2& 231.3216650&USB  & 23& 512 & 0.26 & 5.1 $\times$ 5.0
\tablenotemark{b}& 10.4\\
SO 5$_{6}$--4$_{5}$& 219.9494420&LSB  & 18& 256 & 0.53 & 8.0 $\times$ 6.3
\tablenotemark{b,c}& 68.3\\
H$_2^{13}$CO $3_{12}$--$2_{11}$ &219.9085250&LSB  & 19& 256 & 0.53 & 8.0
$\times$ 6.3\tablenotemark{b,c}& 68.3\\
\enddata
\tablenotetext{a}{
All line frequencies listed were obtained from Pickett et al. (1998)
}
\tablenotetext{b}{
Beam attained after tapering with a 4\as\ $\times$ 4\as\ FWHM Gaussian
}
\tablenotetext{c}{
Beam attained after tapering with a 6\as\ $\times$ 6\as\ FWHM Gaussian
}
\end{deluxetable}

%%%%%%%%%%%%%%%%%% Table 4 %%%%%%%%%%%%%%%%%%%%%%%%%
\begin{deluxetable}{cccccccc}
%\rotate
\tablecolumns{3}
\tablewidth{0pc}
\tablecaption{SMA 1.3 mm peak intensities and masses}
\tablehead{
\colhead{Source} & \colhead{Peak Intensity, I($\lambda$) 
(Jy beam$^{-1}$)\tablenotemark{a}}
& \colhead{Mass (\msun\  beam$^{-1}$)\tablenotemark{b}}}
\startdata
IRS1  & 0.035 & 0.031 \\
IRS2  & 0.023  & 0.020\\
SMA-N  & 0.023 & 0.020 (0.055)\tablenotemark{c}\\
SMA-S  & 0.029  & 0.026 (0.070)\tablenotemark{c}
\enddata

\tablenotetext{a}{Uncertainties on the peak intensities are ~4 mJy beam$^{-1}$.}
\tablenotetext{b}{Masses have been calculated based on the equation,\\
$Mass~ (\msun~beam^{-1})=\frac{I(\lambda)~d^2}{\kappa_{\lambda} B_{\lambda}
(T_{dust})} =0.88 \times I(\lambda)~ (Jy~beam^{-1})$ at $d=300$ pc and 
$\lambda=1.3$ mm, which is derived from
the assumptions of (1) the 1.3 mm opacity ($\kappa_{\lambda}$) of $2\times 
10^{-2}$ cm$^2$ g$^{-1}$ (Kruegel \& Siebenmorgen 1994) and 
(2) the dust temperature ($T_{dust}$) of 20 K for the Planck function, 
$B_{\lambda}$.}
\tablenotetext{c}{The mass if $T_{dust}=10$ K.}
\end{deluxetable}

%%%%%%%%%%%%%%%%%% Table 5 %%%%%%%%%%%%%%%%%%%%%%%%%
\begin{deluxetable}{ccc}
%\rotate
\tablecolumns{3}
\tablewidth{0pc}
\tablecaption{Integrated Intensities of Molecular Lines (K km s$^{-1}$)}
\tablehead{
\colhead{Line} & \colhead{L1251B} & \colhead{East Core} 
}
\startdata
HCO$^+$ 1$-$0  & 6.6 & 7.6  \\
H$^{13}$CO$^+$ 1$-$0  & 1.3 & 1.1\\
CS 2$-$1  & 3.2  & 2.8 \\
HCN 1$-$0 & 4.2 (4)\tablenotemark{a} & 3.0 (0.85)\tablenotemark{a} \\
N$_2$H$^+$ 1$-$0 & 7.8 (2.2)\tablenotemark{a} & 3.8 (1.7)\tablenotemark{a} 
\enddata

\tablenotetext{a}{The optical depth of the main component obtained by the 
hyperfine structure fit.
}
\end{deluxetable}

\clearpage

\begin{figure}
\figurenum{1}
\epsscale{1.0}
\vspace*{-3.0cm}
\plotone{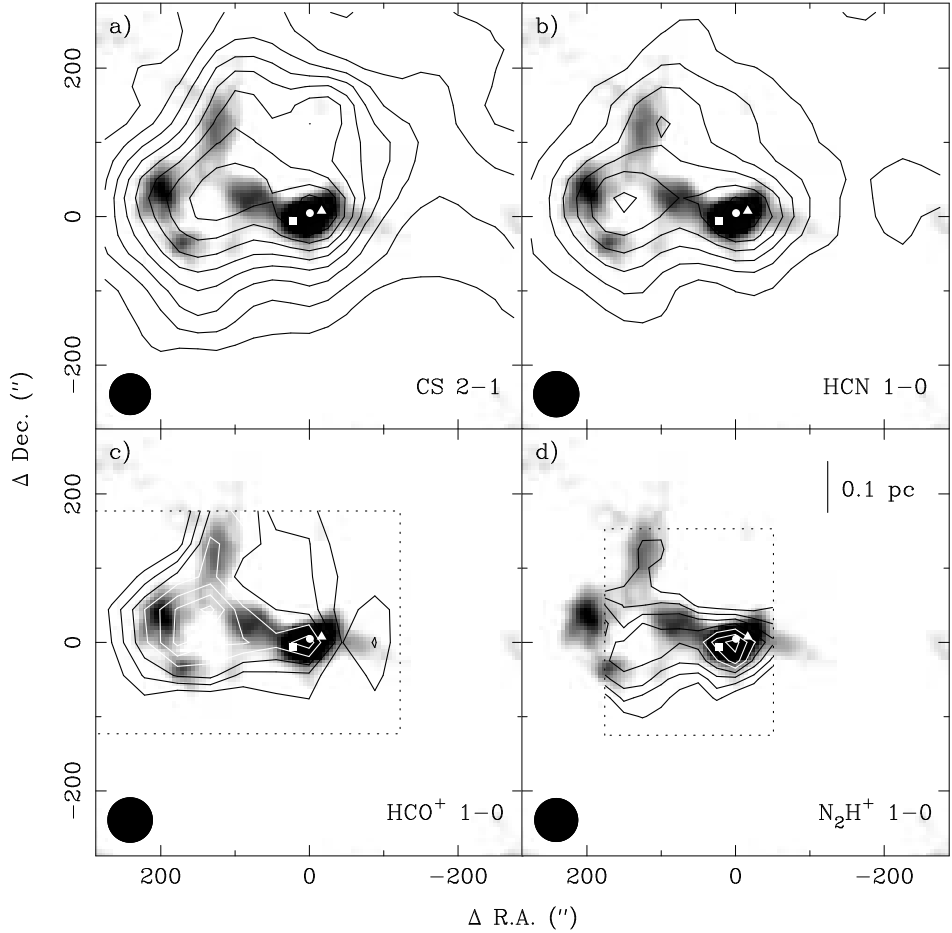}
\vskip -6.0cm
\caption{
Integrated intensity maps (contours) of CS 2$-1$, HCN 1$-$0,
HCO$^+$ 1$-0$, and N$_2$H$^{+}$ 1$-0$ on top of the 850 \micron\
emission map (grey scale).  The (0,0) position of this map represents
the coordinates of IRAS 22376+7455.  In each panel, the small white
circle, box and triangle denote the positions of IRS1, IRS2 and IRS4
respectively.  In each panel, the grey scale used ranges from 0.084
to 0.420 Jy beam$^{-1}$, roughly 2 $\sigma$ to 10 $\sigma$, where 1
$\sigma$ $\approx$ 0.042 Jy beam$^{-1}$.  In each panel, the large
black circle at lower left shows the beam FWHM.  In each panel, the
dotted rectangle shows the extent of the map in the respective line.
(For CS 2$-$1 and HCN 1$-$0, the map was larger than the area shown.)
a) Integrated intensity map of CS 2$-$1.  The velocity range used is
$-6.5$ to $-0.5$ km s$^{-1}$.  The contours begin at 10 $\sigma$ and
increase by 6 $\sigma$ where 1 $\sigma = 0.06$ K km s$^{-1}$.  b)
Integrated intensity map of HCN 1$-$0.  The velocity range used
is $-13.5$ to 4 km s$^{-1}$.  The contours begin at 10 $\sigma$ and
increase in steps of 5 $\sigma$ where 1 $\sigma$ = 0.106 K km s$^{-1}$.
c) Integrated intensity map of HCO$^+$ 1$-0$.  The velocity range used
is $-6$ to $-2$ km s$^{-1}$.  The contours begin at 6 $\sigma$ and
increase in steps of 1.5 $\sigma$ where 1 $\sigma$ = 0.42 K km s$^{-1}$.
d) Integrated intensity map of N$_2$H$^{+}$ 1$-0$.  The velocity range
used is $-13$ to 4 km s$^{-1}$.  The contours begin at 15 $\sigma$ and
increase in steps of 5 $\sigma$ where 1 $\sigma$ = 0.166 K km s$^{-1}$.
}
\end{figure}

\clearpage

\begin{figure}
\figurenum{2}
\vspace*{-3cm}
\epsscale{1.0}
\plotone{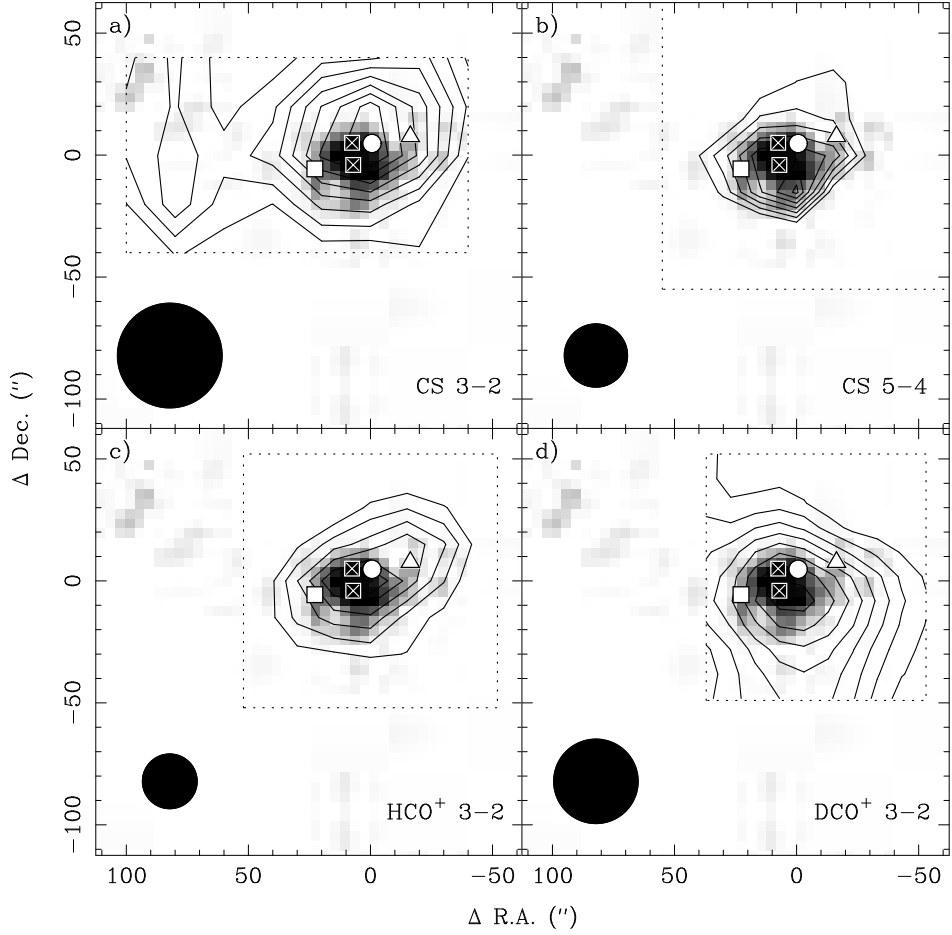}
\vskip -6cm
\caption{
Integrated intensity maps (contours) of CS $3-2$, $5-4$, HCO$^+$ 3$-$2,
and DCO$^+$ 3$-$2 on top of the 1.3 mm dust continuum emission map (grey
scale).  The grey scale ranges from 0.025 to 0.125 Jy beam$^{-1}$ or
about 2 $\sigma$ to 10 $\sigma$, where 1 $\sigma$ = 0.0125 Jy beam$^{-1}$.
In each panel, the dotted rectangle denotes the extent of the map in the
respective line.  Symbols are defined as in Figure 1 but here the square-X
symbols indicate positions of peak intensity of prestellar condensations
detected in SMA 1.3 mm observations (see Figure 8 of Paper I).  a) Integrated
intensity map of CS 3$-$2.  The velocity range used is $-6.5$ to $-0.5$ km
s$^{-1}$.  The contours begin at 15 $\sigma$ and increase in steps of 5
$\sigma$ where 1 $\sigma=0.066$ K km s$^{-1}$.  b) Integrated intensity map
of CS 5$-$4.  The velocity range used is $-6.5$ to $-0.5$ km s$^{-1}$.  The
contours begin at 10 $\sigma$ and increase in steps of 4 $\sigma$ where 1
$\sigma = 0.112$ K km s$^{-1}$.  c) Integrated intensity map of HCO$^+$ 3$-$2.
The velocity range used is $-8$ to $0$ km s$^{-1}$.  The contours begin at
3 $\sigma$ and increase in steps of 3 $\sigma$ where 1 $\sigma = 0.545$ K
km s$^{-1}$.  d) Integrated intensity map of DCO$^+$ 3$-$2.  The velocity
range used is $-5$ to $-2.5$ km s$^{-1}$.  The contours begin at 3 $\sigma$
and increase in steps of 1 $\sigma$ where 1 $\sigma$ = 0.097 K km s$^{-1}$.
}
\end{figure}

\clearpage

\begin{figure}
\figurenum{3}
\epsscale{1.0}
\plotone{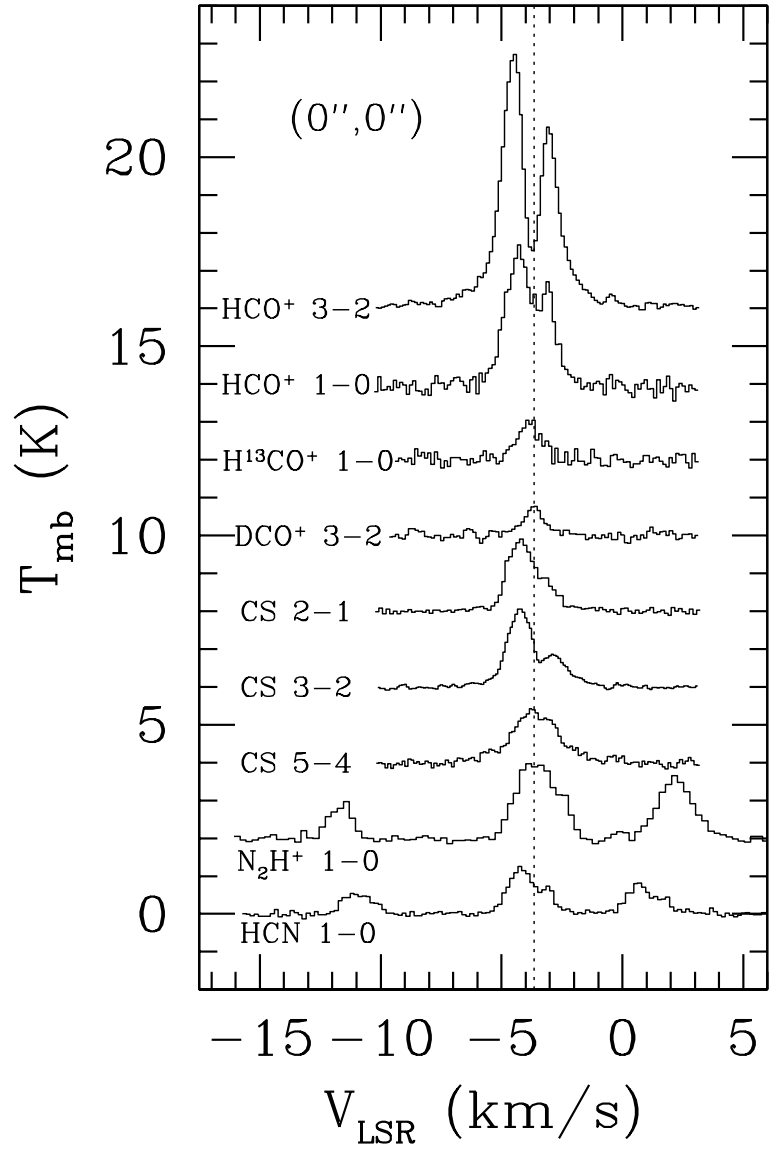}
\caption{
The spectra of all molecular transitions observed with single dish telescopes 
(Table 1) at the center position. 
To present line profiles better, the order of lines has been 
chosen arbitrarily, and spectra have been shifted up. 
The dotted vertical line indicates the centroid velocity of L1251B, -3.65 \kms.
}
\end{figure}

\clearpage

\begin{figure}
\figurenum{4}
\plotone{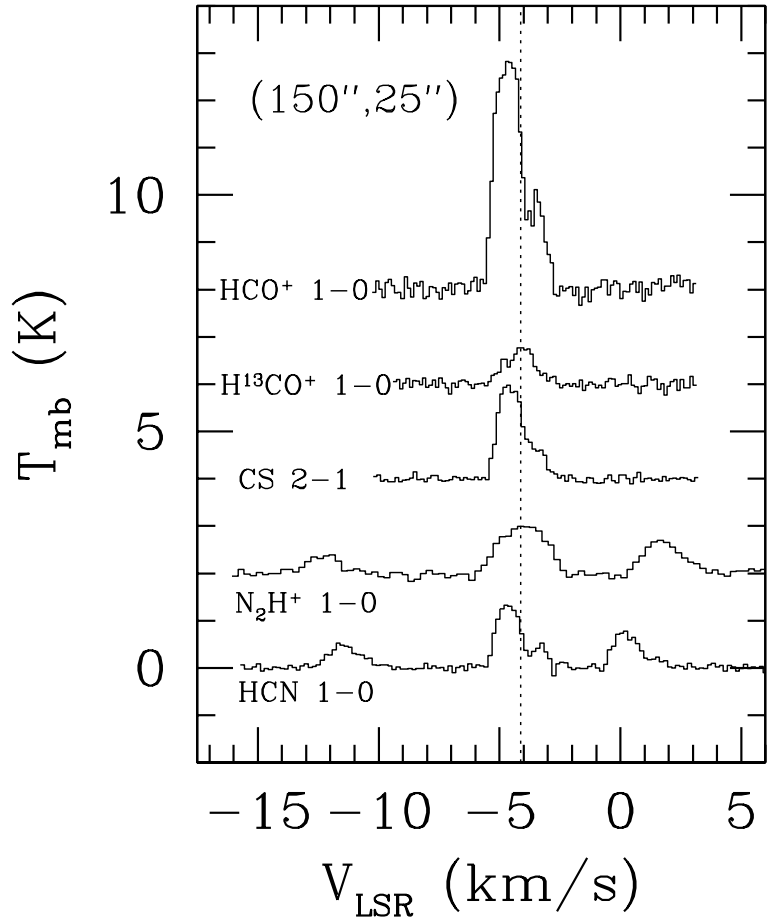}
\caption{
The spectra of some molecular transitions observed with single-dish telescopes 
toward the starless east core at (150$\as$, 25$\as$). 
To present line profiles better, the order of lines has been
chosen arbitrarily, and spectra have been shifted up.
The dotted vertical line indicates the centroid velocity of the east core, 
-4.13 \kms\, which is derived from the Gaussian line fitting of H$^{13}$CO$^+$ 
1$-$0.
}
\end{figure}

\clearpage

\begin{figure}
\figurenum{5}
\vspace*{-4.0cm}
\epsscale{0.6}
\plotone{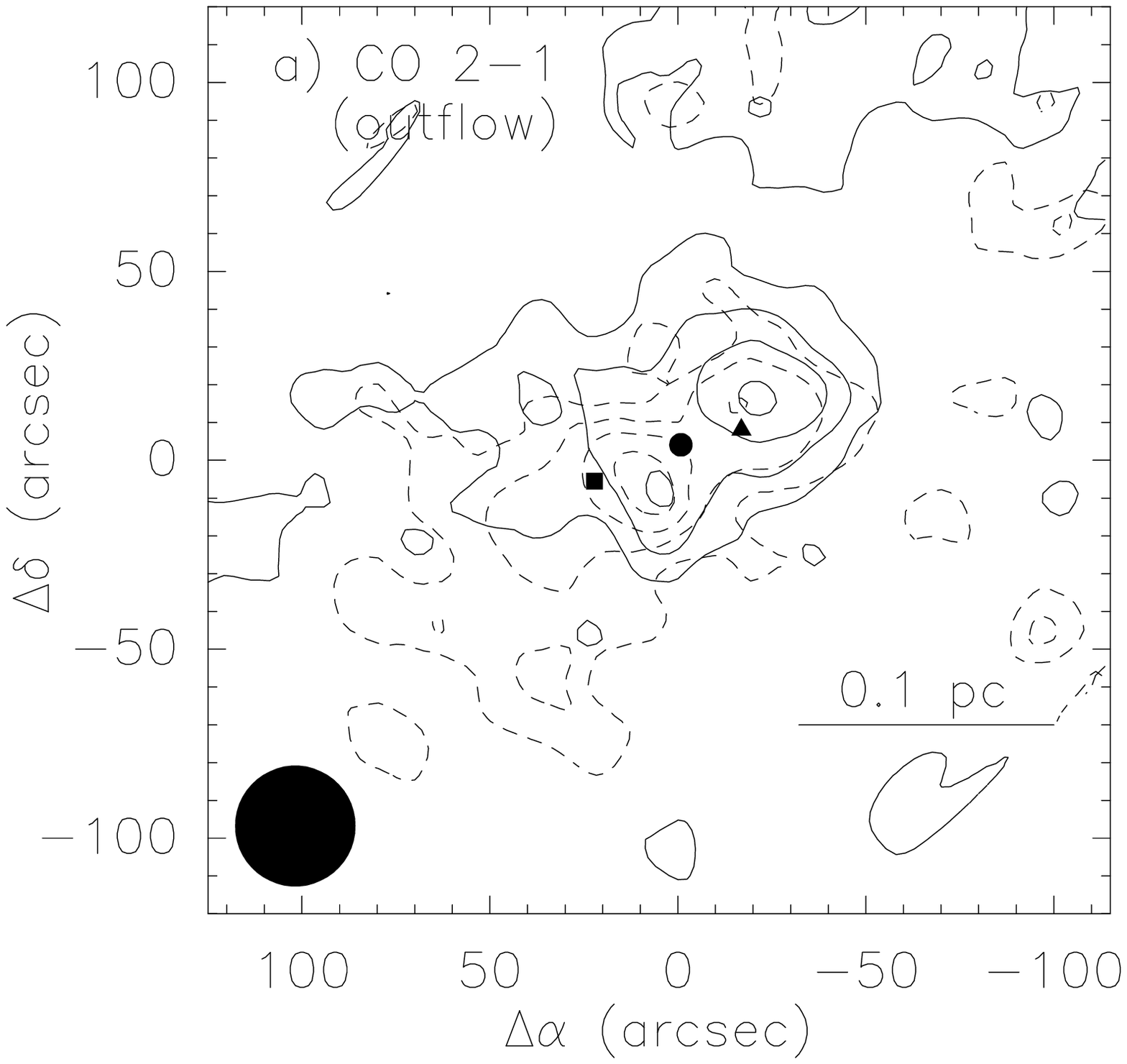}
\vspace*{-0.5cm}
\plotone{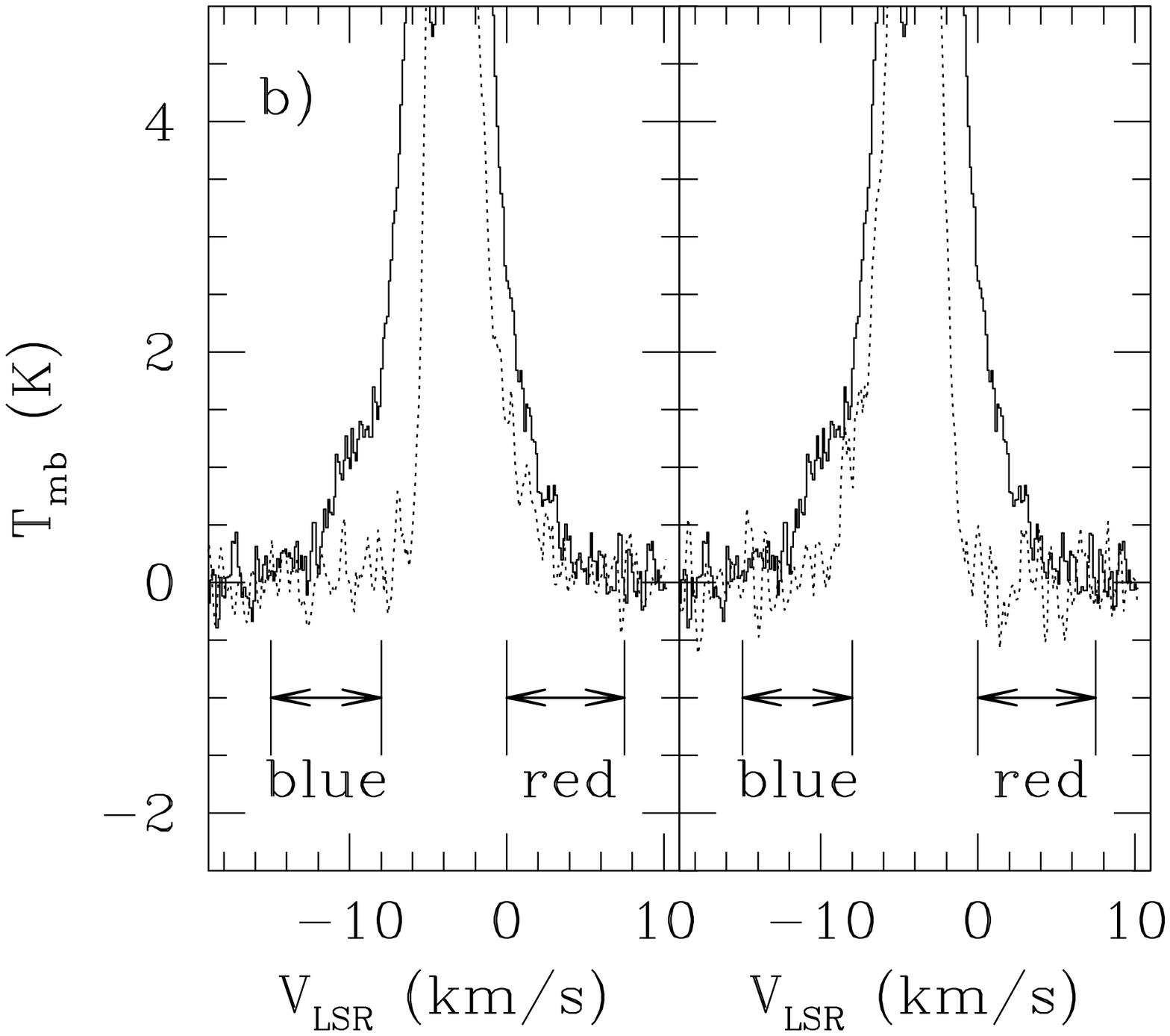}
\vspace*{-1cm}
\caption{
a) The distribution of the outflow in L1251B in CO 2$-$1.  
Symbols are defined as in Figure 1.  
The solid contours and dashed contours represent the blue and 
red components of the outflow, respectively. Contours are every 
1 K km s$^{-1}$ from 1 K km s$^{-1}$ (1 $\sigma$ = 0.5 K km s$^{-1}$).  
The velocity range used for the blue and red components 
are $-15$ to $-8$ km s$^{-1}$ and 0 to 7.5 km s$^{-1}$, respectively.
b) The spectrum at the center compared with blue- (left) and red-free (right)
spectra. Arrows indicate the velocity ranges for the blue and red components. 
}
\end{figure}

\clearpage

\begin{figure}
\figurenum{6}
\centering
 \vspace*{7.5cm}
   \leavevmode
  \includegraphics{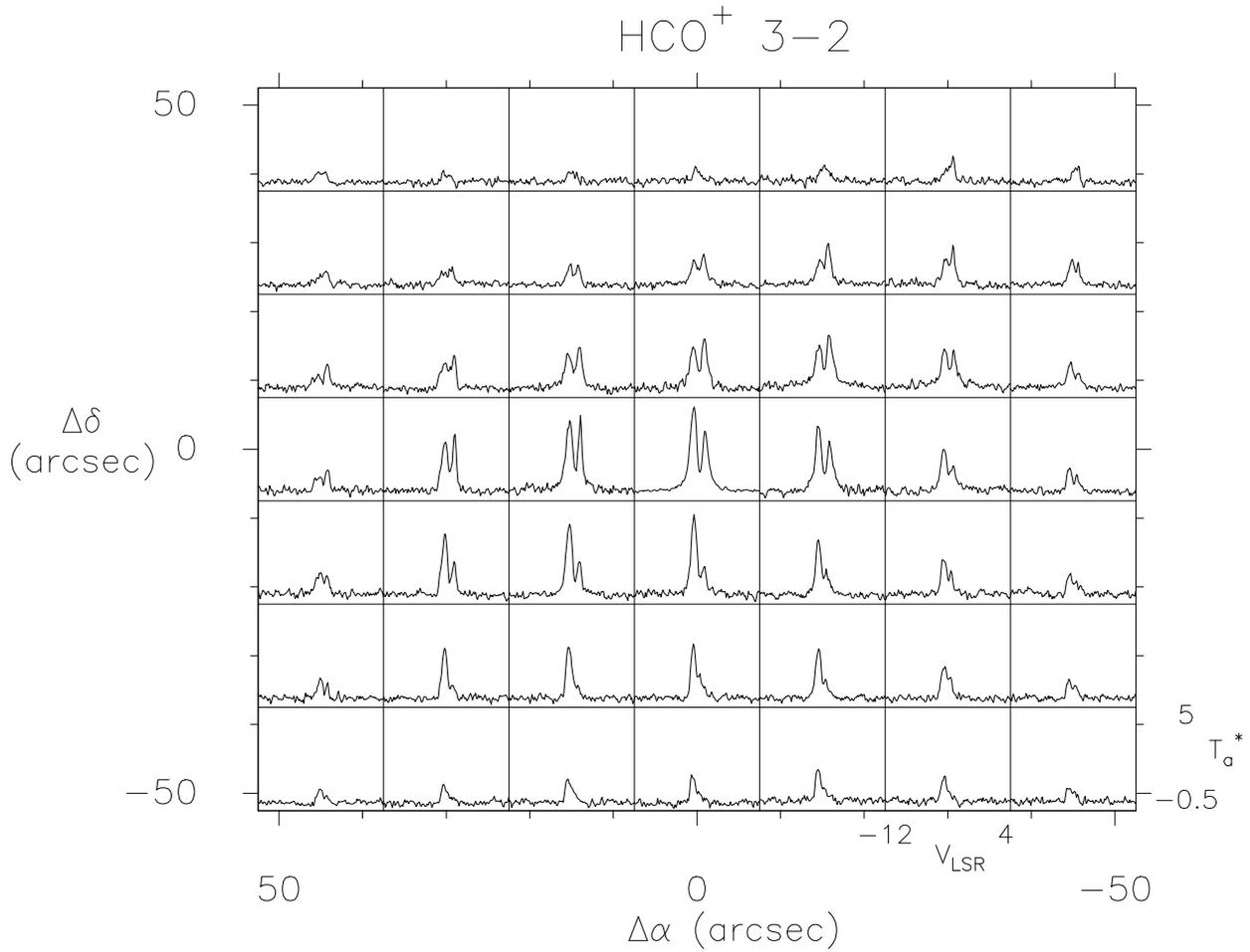}
\vskip 3.00in
\figcaption{
The spectra of HCO$^+$ $3-2$ across L1251B.  Panels are spaced by 15$\as$ 
in R.A. and Dec.  The horizontal axis is for $V_{LSR}$ whose range is $-12$ 
km s$^{-1}$ to 4 km s$^{-1}$, and the vertical axis is in T$_a^*$ whose 
range is $-0.5$ K to 5 K. }
\end{figure}

\clearpage

\begin{figure}
\figurenum{7}
\epsscale{0.5}
\vspace*{-0.5cm}
\plotone{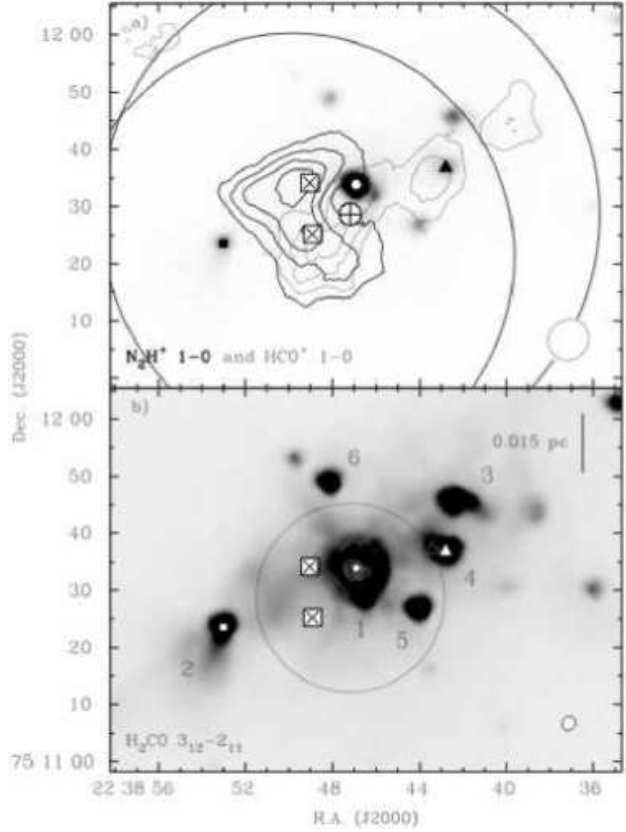}
\vskip -0.0cm
\caption{
Integrated intensities of molecular transitions observed at OVRO, found
by determining the ``zeroth-moment," i.e., summing over channels with
emission $>\mid 2\mid$ $\sigma$.  In all cases, no corrections for primary
beam attenuation have been made.  Each contour set begins at the respective
2 $\sigma$ level and increases in steps of the respective 2 $\sigma$.
Integrated intensities are overlaid onto an image of the same region as
seen in IRAC Band 2 (inverted grey scale).  In each panel, the ellipse to
the lower right denotes the size and P.A. of the synthesized beam FHWM.
Symbols are defined
as in Figure 2, although the position of IRAS 22376+7455 is shown as
a circle-cross symbol.  a) N$_{2}$H$^{+}$ 1$-$0 (thick contours) and
HCO$^{+}$ 1$-$0 (thin contours) where 1 $\sigma$ = 1.5 K km s$^{-1}$ for
N$_{2}$H$^{+}$ 1$-$0 and 1 $\sigma$ = 1.3 K km s$^{-1}$ for HCO$^{+}$
1$-$0.  (The beam FWHM shown is that of the HCO$^{+}$ 1$-$0 data.)  The
outer large circle shows the FWHM of the primary beam of the HCO$^{+}$
data while the inner large circle shows the FWHM of the primary beam of
the N$_{2}$H$^{+}$ data.  For the IRAC Band 2 data, the grey scale range
is -0.5$-$100 MJy sr$^{-1}$.  b) Integrated intensities of H$_{2}$CO
3$_{12}$--2$_{11}$ (contours), where 1 $\sigma$ = 0.54 K km s$^{-1}$.  The
large circle shows the FWHM of the primary beam of H$_{2}$CO data.  For the
IRAC Band 2 data, the grey scale range is -0.5$-$10 MJy sr$^{-1}$.  The
numbers shown identify individual IR sources described in the text.  (IRS1,
IRS2, and IRS4 were classified as Class 0/I objects, and IRS3, IRS5, and
IRS6 were classified as Class II sources in Paper I.)
}
\end{figure}

\clearpage

\begin{figure}
\figurenum{8}
\vspace*{-4cm}
\epsscale{1.0}
\plotone{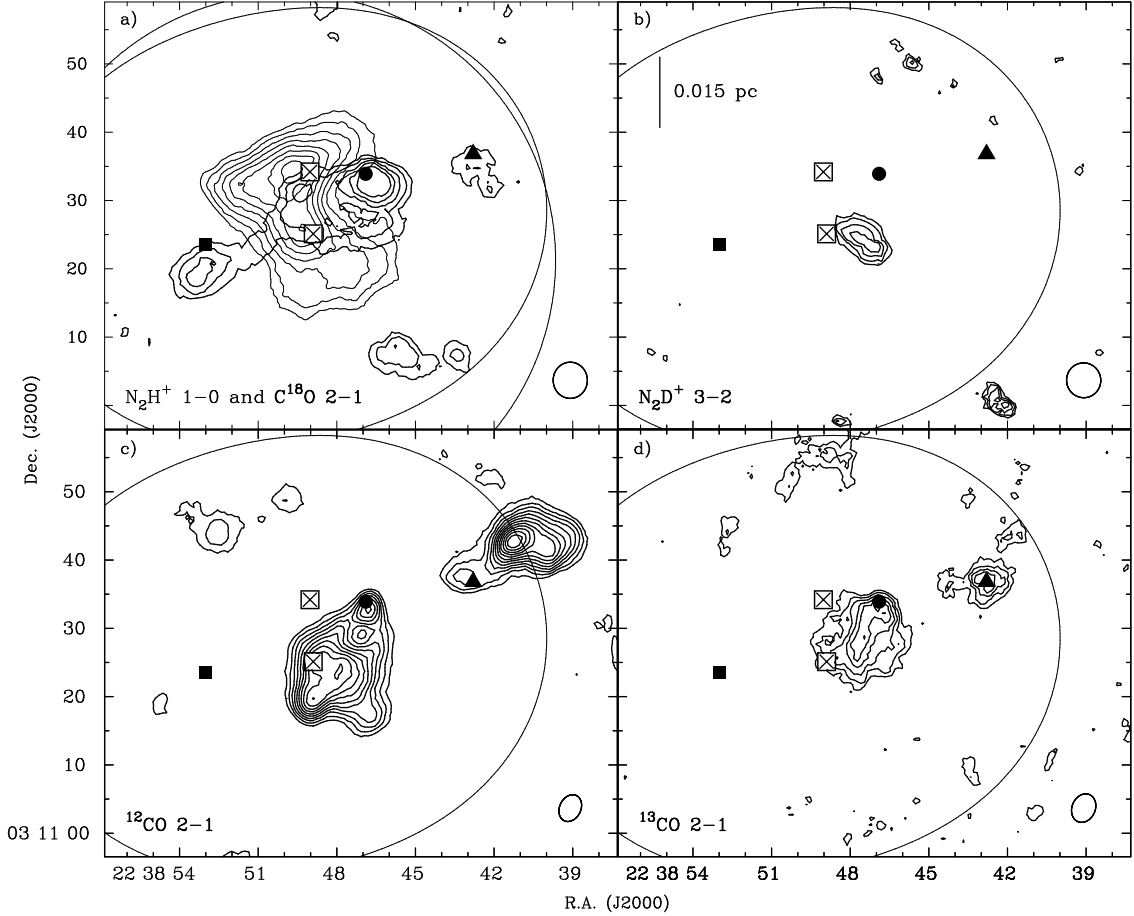}
\vskip -5.5cm
\caption{
Integrated intensities of molecular transitions observed at SMA,
found by determining the ``zeroth-moment," i.e., summing over channels
with emission $>$ $\mid 2\mid$ $\sigma$.  Each contour set begins at
the respective 2 $\sigma$ level and increases in steps of respective
1 $\sigma$.  In each panel, the ellipse to the lower right denotes
the size and P.A. of the respective synthesized beam FWHM.  In each
panel, the large oval denotes the extent of the half-power sensitivity
of the two-pointing mosaic, obtained from the continuum data obtained
when the lines were observed.  (The primary beam attenuation was
corrected when these mosaics were made.)  Various symbols are defined
as in previous Figures.  a) Integrated intensities of C$^{18}$O 2$-$1
(thick contours) where 1 $\sigma$ $\approx$ 0.5 K km s$^{-1}$.  For
comparison, the integrated intensities of N$_{2}$H$^{+}$ 1$-$0 from
Figure 7a are also shown.  (The outer large circle denotes the primary
beam FWHM of the N$_{2}$H$^{+}$ data.)  b) Integrated intensities of
N$_{2}$D$^{+}$ 3$-$2 where 1 $\sigma$ $\approx$ 0.5 K km s$^{-1}$.
(The contours just south of the half-power sensitivity oval are tracing
amplified noise and not N$_{2}$D$^{+}$ emission.)  Integrated intensities
of $^{12}$CO 2$-$1 where 1 $\sigma$ $\approx$ 7 K km s$^{-1}$.  d)
Integrated intensities of $^{13}$CO 2$-$1 where 1 $\sigma$ $\approx$
2 K km s$^{-1}$.
}
\end{figure}

\clearpage

\begin{figure}
\figurenum{9}
\vspace*{-2cm}
\epsscale{1.0}
\plotone{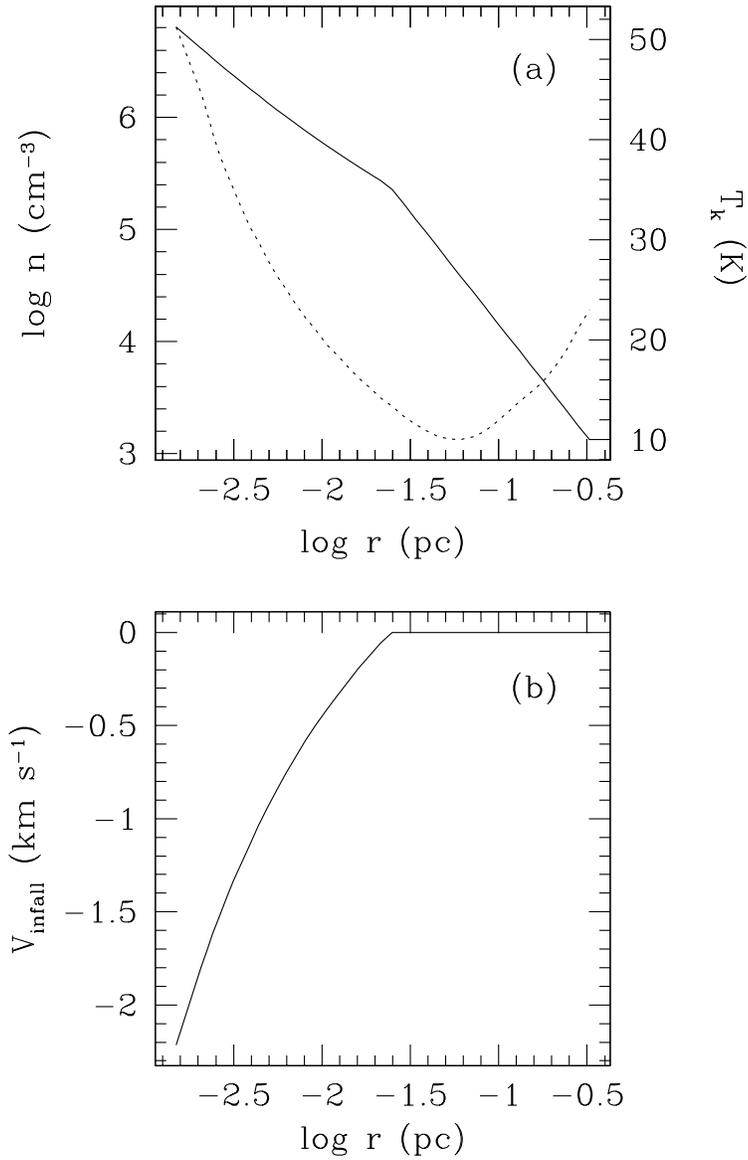}
\caption{
Density (a) and infall velocity (b) profiles of the best-fit
inside-out collapse model of L1251B for the dust continuum observations
at 450 \micron\ and 850 \micron.  The infall radius is 5000 AU, equivalent 
to an infall timescale of $5\times 10^4$ years.  The kinetic temperature 
profile in (a) has been calculated by balancing the heating and cooling 
of gas from the dust temperature profile obtained from the dust continuum 
modeling (See Young et al. 2003 for details of the dust modeling process.)
At small radii, the kinetic temperature is well coupled with the dust 
temperature, but at large radii, it is higher than the dust temperature 
mainly due to the photoelectric heating.
}
\end{figure}

\clearpage

\begin{figure}
\figurenum{10}
\epsscale{0.5}
\vspace*{-2.0cm}
\plotone{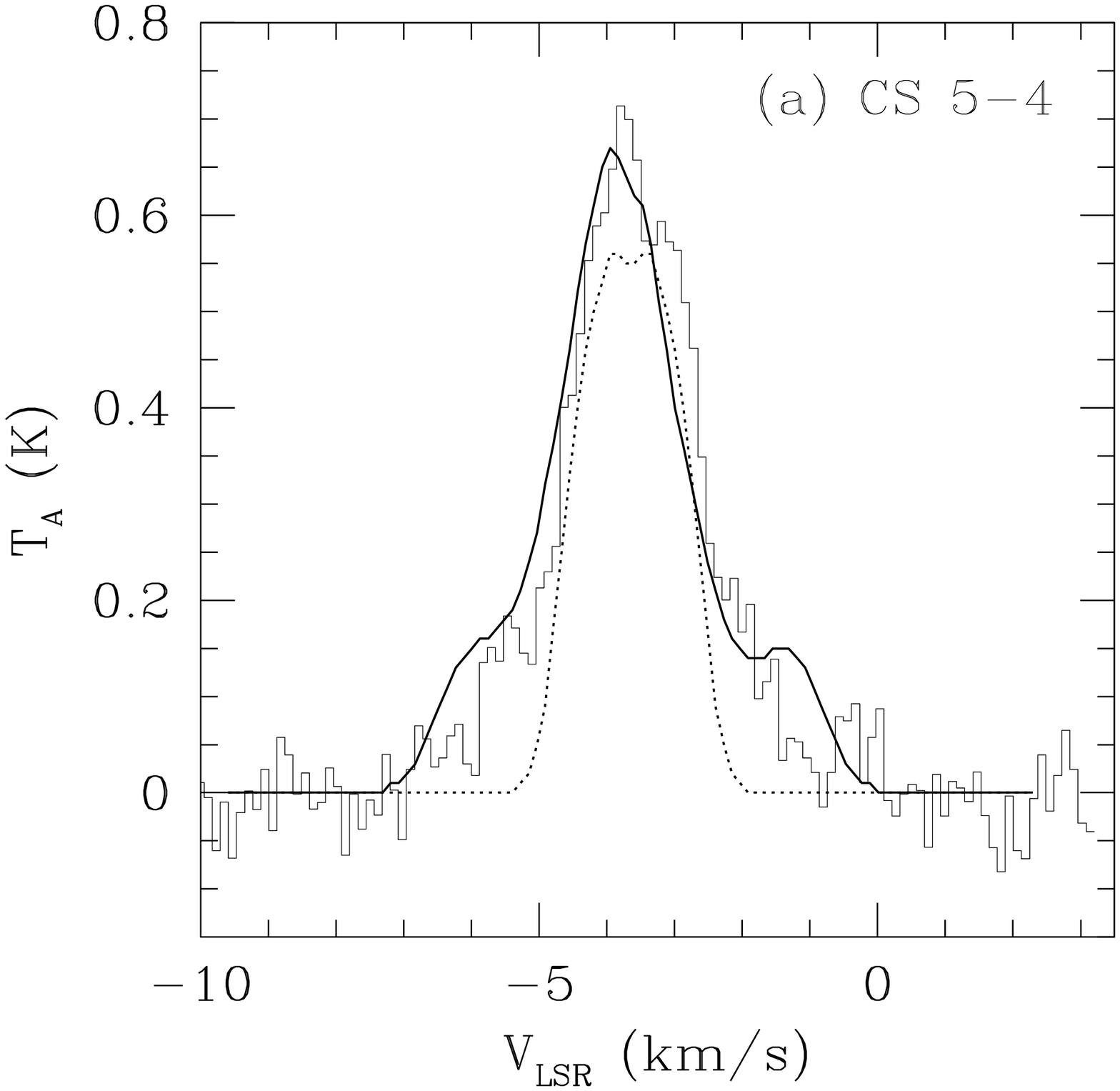}
\vspace*{-0.5cm}
\plotone{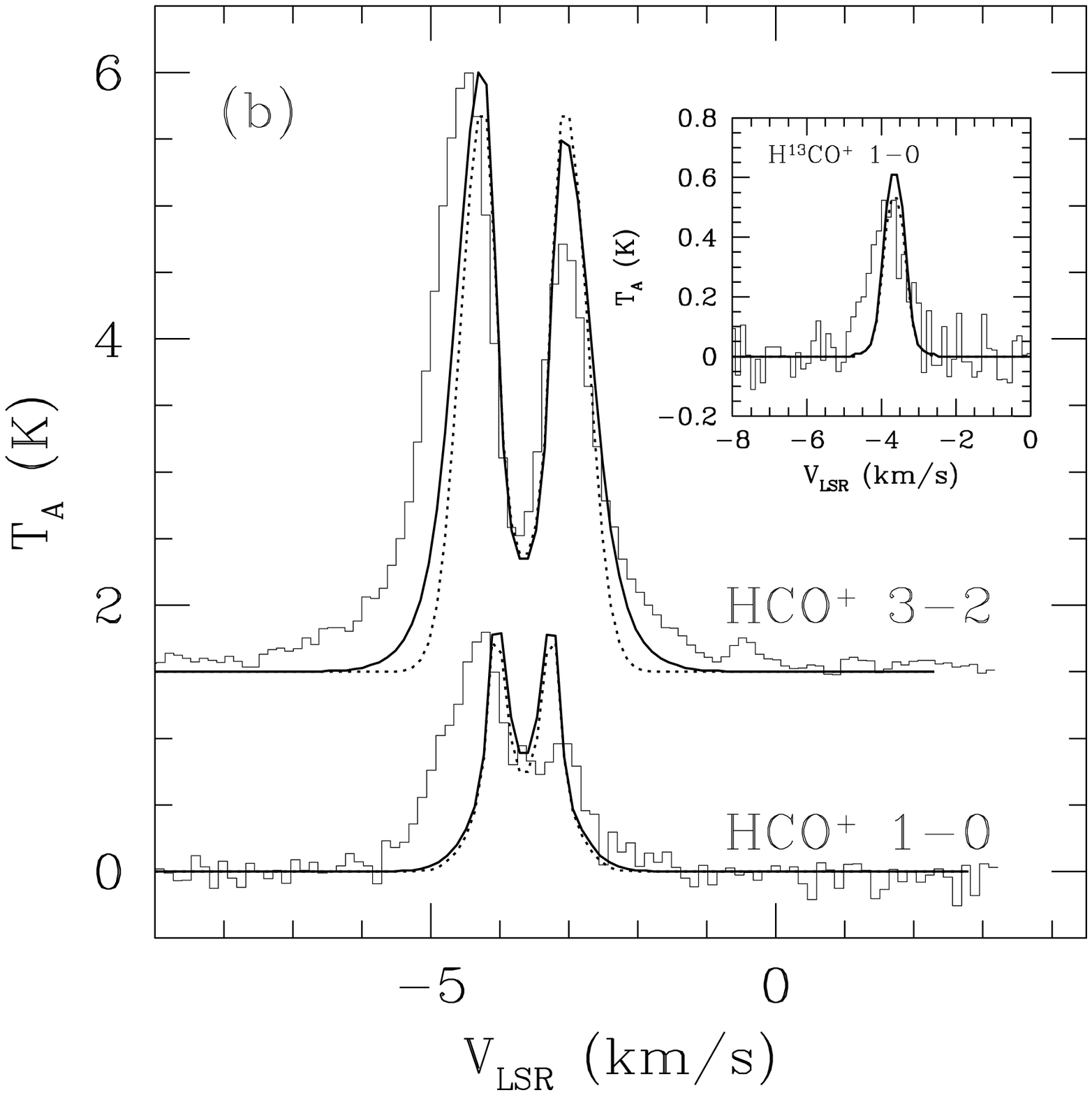}
\vspace*{-1.0cm}
\caption{
Comparisons between observations (histogram) and models with (solid line) 
and without (dotted line) the consideration of infall velocity profile
in CS 5$-$4 (a) and HCO$^+$ 3$-$2 and H$^{13}$CO$^+$ 1$-$0 (b).
For the line radiative transfer calculations, physical conditions in Figure 9 
have been used, and the abundance profiles were adopted from Fig. 18. 
For the H$^{13}$CO$^+$ abundance profile, the abundance ratio, 
$^{12}$C/$^{13}$C$=77$ was assumed, and the HCO$^+$ abundance profile for
HCO$^+$ 3$-$2 has been reduced by a factor of 2 from the profile seen in 
Fig. 18.  The infall velocity profile is important to produce the broad wings 
of CS 5$-$4, which traces very high densities. 
The line wings of HCO$^+$ 3$-$2 and H$^{13}$CO$^+$ 1$-$0 may be also 
affected by outflow, which is not considered in this modeling. }
\end{figure}

\clearpage

\begin{figure}
\figurenum{11}
\epsscale{1.0}
\plotone{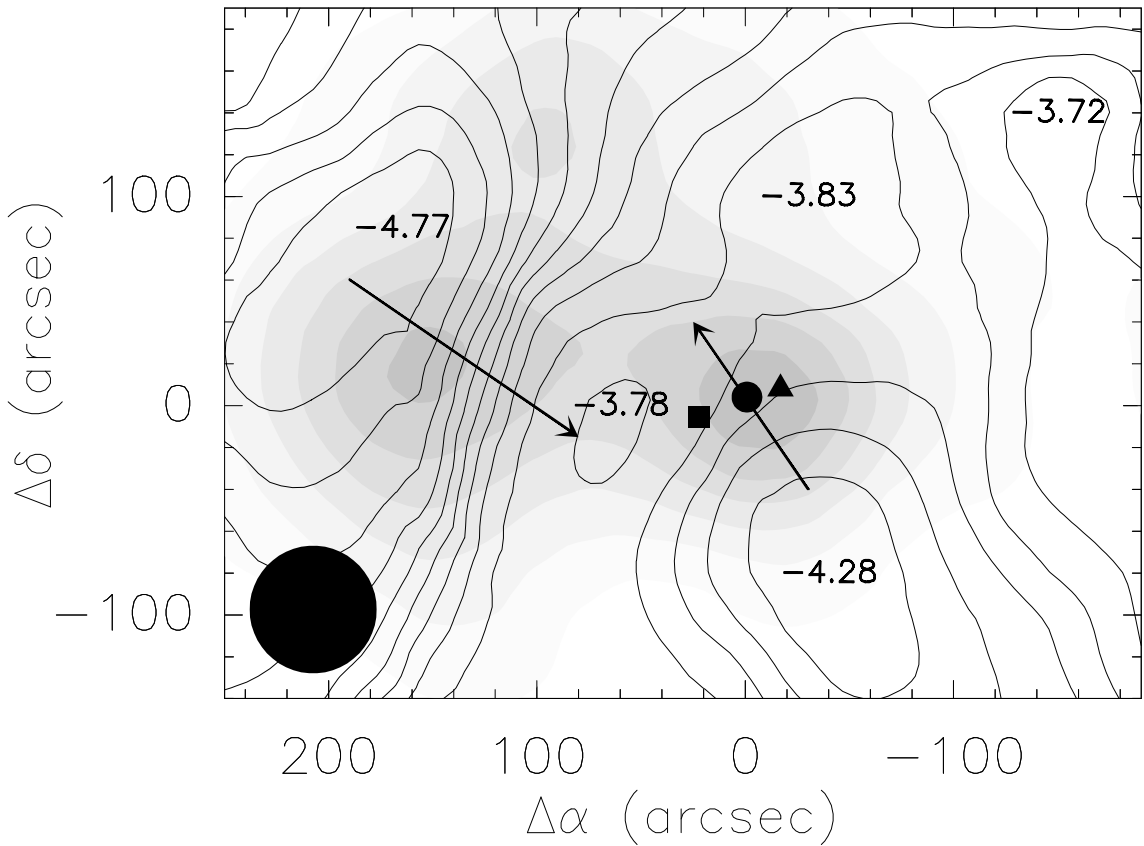}
\vskip -8cm
\caption{
The centroid velocity distribution of HCN $1-0$ as observed from FCRAO
overlaid onto its integrated intensity map. Centroid velocities were
calculated by fitting the hyperfine structure of HCN $1-0$.  
The interval of contours
is 0.1 km s$^{-1}$ and numbers indicate local extrema of centroid velocity.
The arrows indicate the directions of velocity gradients from blue to red.
The uncertainty of fitting of the centroid velocity is less than 0.1 \kms.
}
\end{figure}

\clearpage

\begin{figure}
\figurenum{12}
\centering
 \vspace*{6.5cm}
   \leavevmode
  \includegraphics{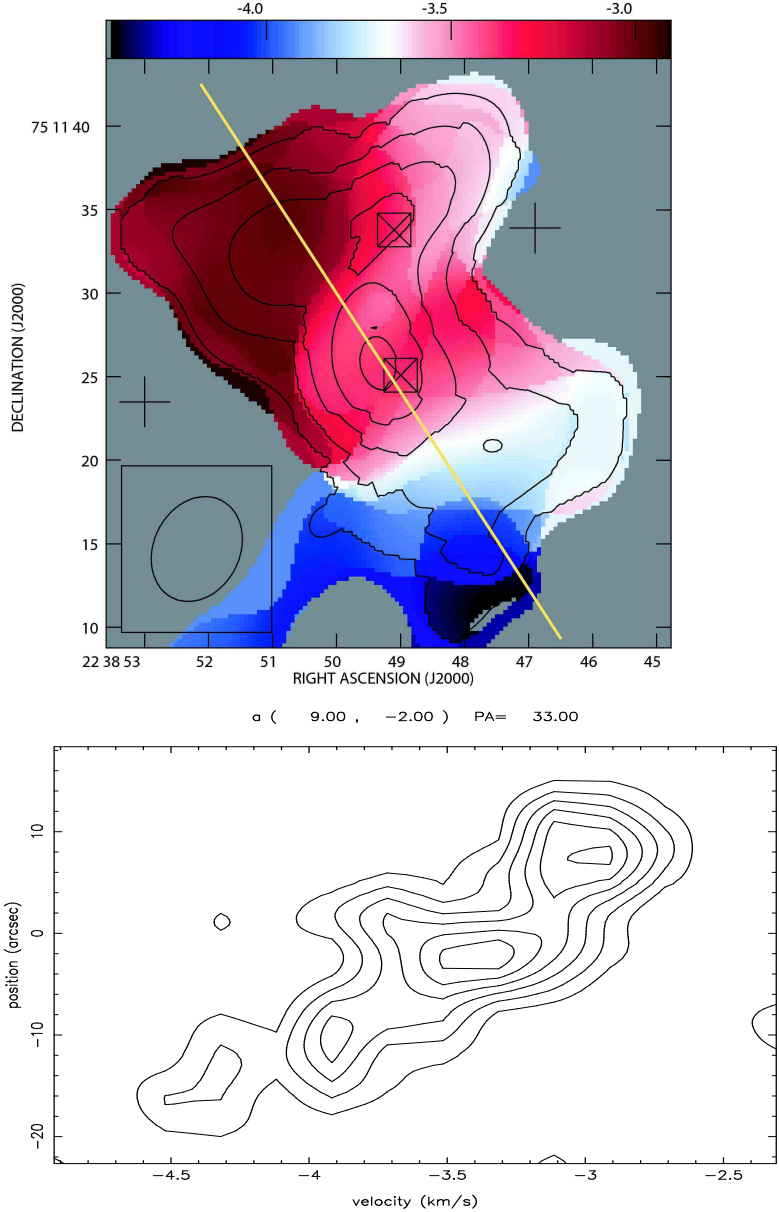}
  \includegraphics{fig12.b.ps}
\vskip 3.60in
\caption{
(a) The integrated intensity contours overlaid onto the mean centroid velocity 
image of the isolated component (F$=01-12$) of N$_2$H$^+$ 1$-$0 observed at
the OVRO.
The moments were calculated in channels with emission $> 2 \sigma$.
The contours begin at 3 $\sigma$ and increase in steps of 3 $\sigma$, where 1 
$\sigma$ is 1.5 K \kms. Crosses denote the positions of IRS1 and IRS2, and 
the square-X symbols indicate the positions of SMA-N and SMA-S.
The ellipse to the lower left represents the synthesized beam of the 
N$_2$H$^+$ 1$-$0 observation. The overlaid line indicates the direction for the 
position-velocity diagram in (b).
(b) The position-velocity diagram of the isolated component along the cut that 
centers at the emission peak,   
$(\Delta \alpha, \Delta \delta)=(9\arcsec,-2\arcsec)$ with P.A.$=33\degree$.
This cut is basically along the integrated intensity map elongated in the
direction of northeast--southwest crossing the emission peak. 
For these figures, the isolated component has been shifted in velocity by
8.0 \kms.
The contours begin at 2 $\sigma$ and increase in steps of 2 $\sigma$, where 
1 $\sigma$ is 0.25 K beam$^{-1}$. 
}
\end{figure}

\clearpage

\begin{figure}
\figurenum{13}
\epsscale{1.0}
\vskip -3cm
\plotone{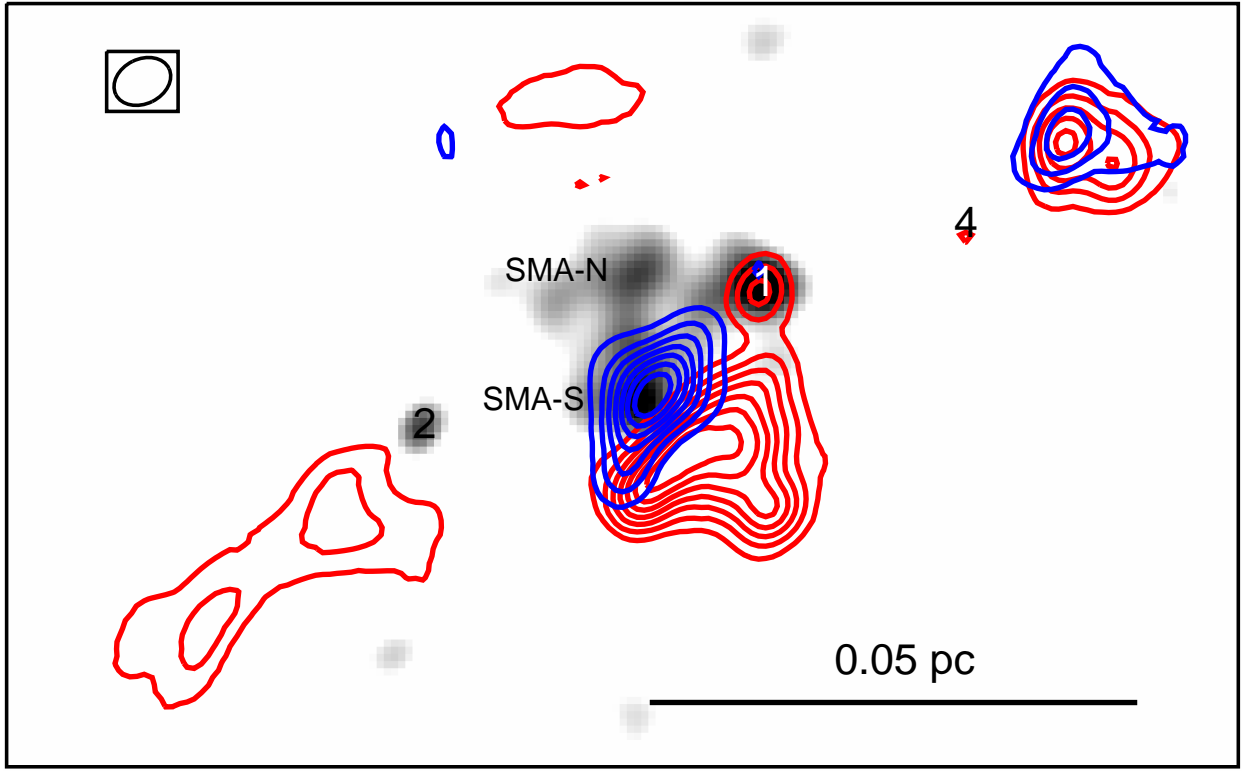}
\caption{
Integrated intensities of red- and blueshifted emission of $^{12}$CO 2$-$1 
overlaid onto the 1.3 mm continuum emission image observed at the SMA.
Red and blue contours denote intensities calculated over 
``blue" and ``red" velocity ranges, respectively.  
The same velocity ranges as in Figure 5 were used. 
Numbers indicate positions of IRS1, IRS2, and IRS4. Two starless condensations
are named SMA-N and SMA-S.  
The ellipse to the upper left denotes the synthesized beam of the 
CO 2$-$1 observation.
The contours begin at 4 $\sigma$ and increase in steps of 4 $\sigma$, 
where 1 $\sigma$ = 2.14 K \kms. 
The grey scale range is $0.01-0.0275$ Jy beam$^{-1}$.
}
\end{figure}

\clearpage

\begin{figure}
\figurenum{14}
\vskip -2cm
\plotone{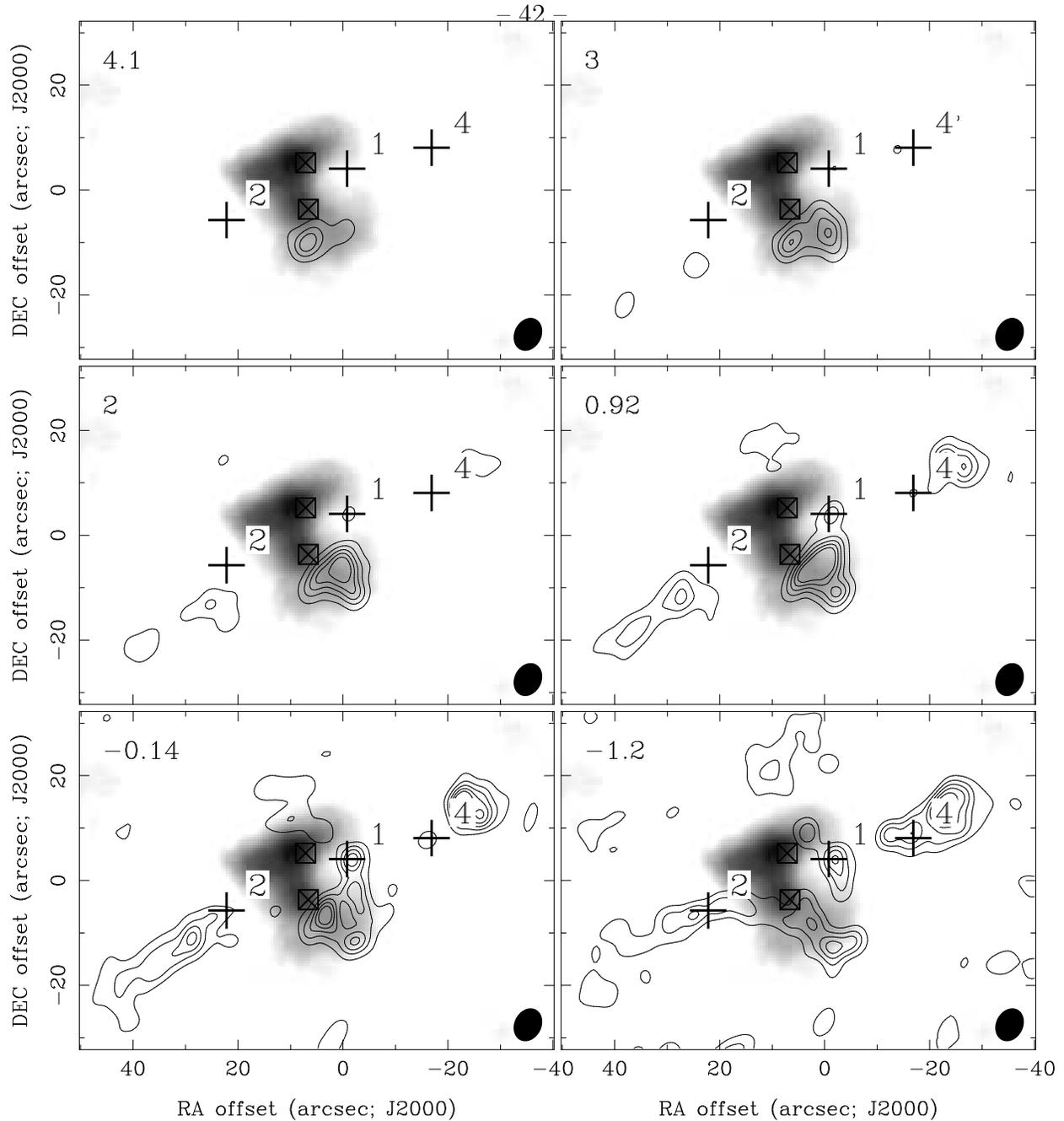}
\caption{
The channel map of redshifted emission of $^{12}$CO 2$-$1 observed at the SMA
overlaid onto the integrated intensity image of N$_2$H$^+$ 1$-$0 observed at
the OVRO.
The contours begin at 6 $\sigma$ and increase in steps of 9 $\sigma$, 
where 1 $\sigma$ is 0.36 K.
The grey scale range is 1.5 K \kms\ to 15 K \kms.
Two square-X symbols indicate the positions of SMA-N and SMA-S, 
and crosses denote the positions of IRS1, IRS2, and IRS4.
The ellipse to the lower right in each panel denotes the synthesized beam of
the CO 2$-$1 observation.
}
\end{figure}

\clearpage

\begin{figure}
\figurenum{15}
\vskip -2cm
\plotone{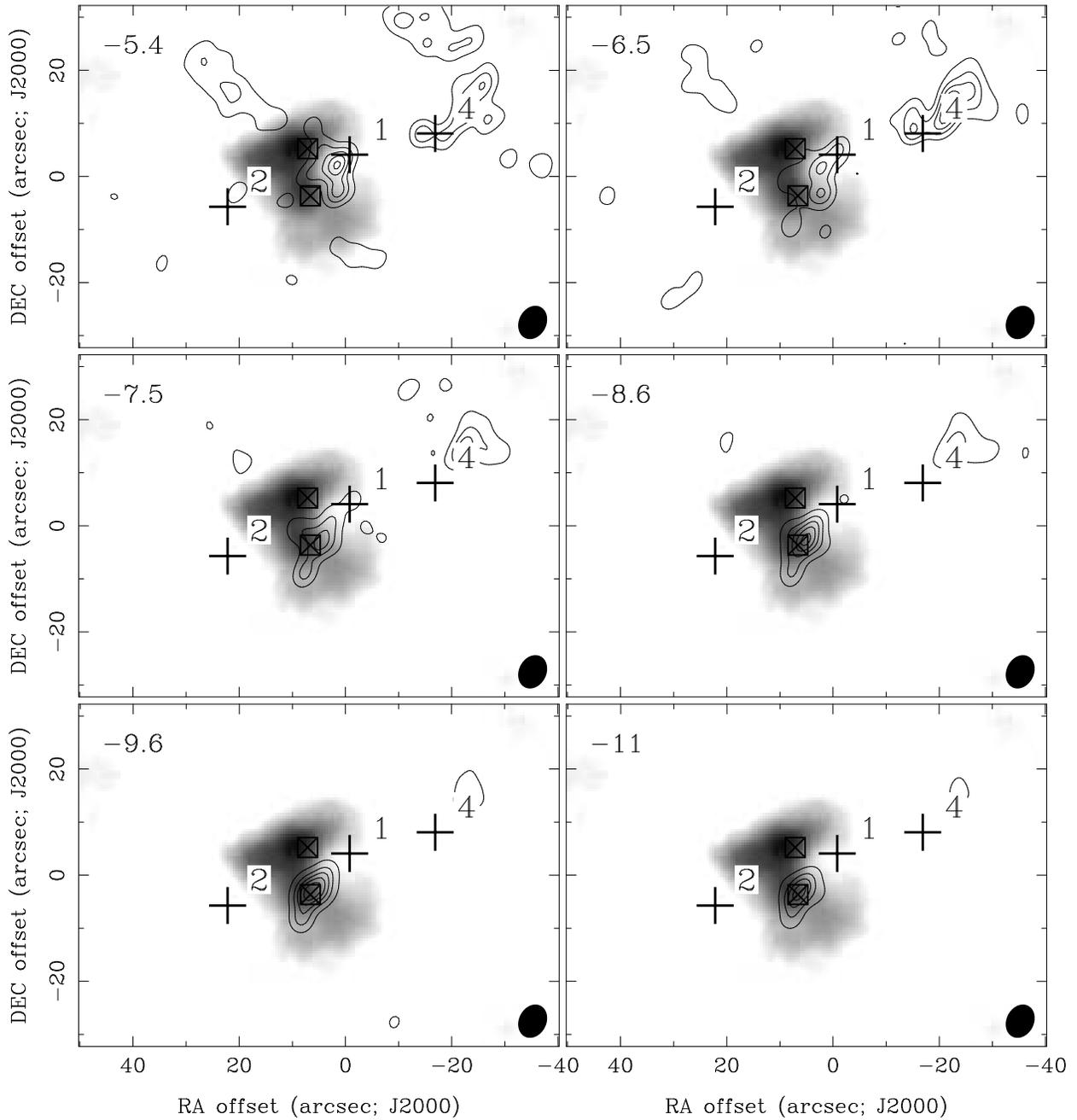}
\caption{
The channel map of blueshifted emission of $^{12}$CO 2$-$1 observed at the SMA
overlaid onto the integrated intensity image of N$_2$H$^+$ 1$-$0 observed at
the OVRO.
The contour levels and grey scale are the same as in Figure 14.
}
\end{figure}

\clearpage

\begin{figure}
\figurenum{16}
\vskip -4cm
\plotone{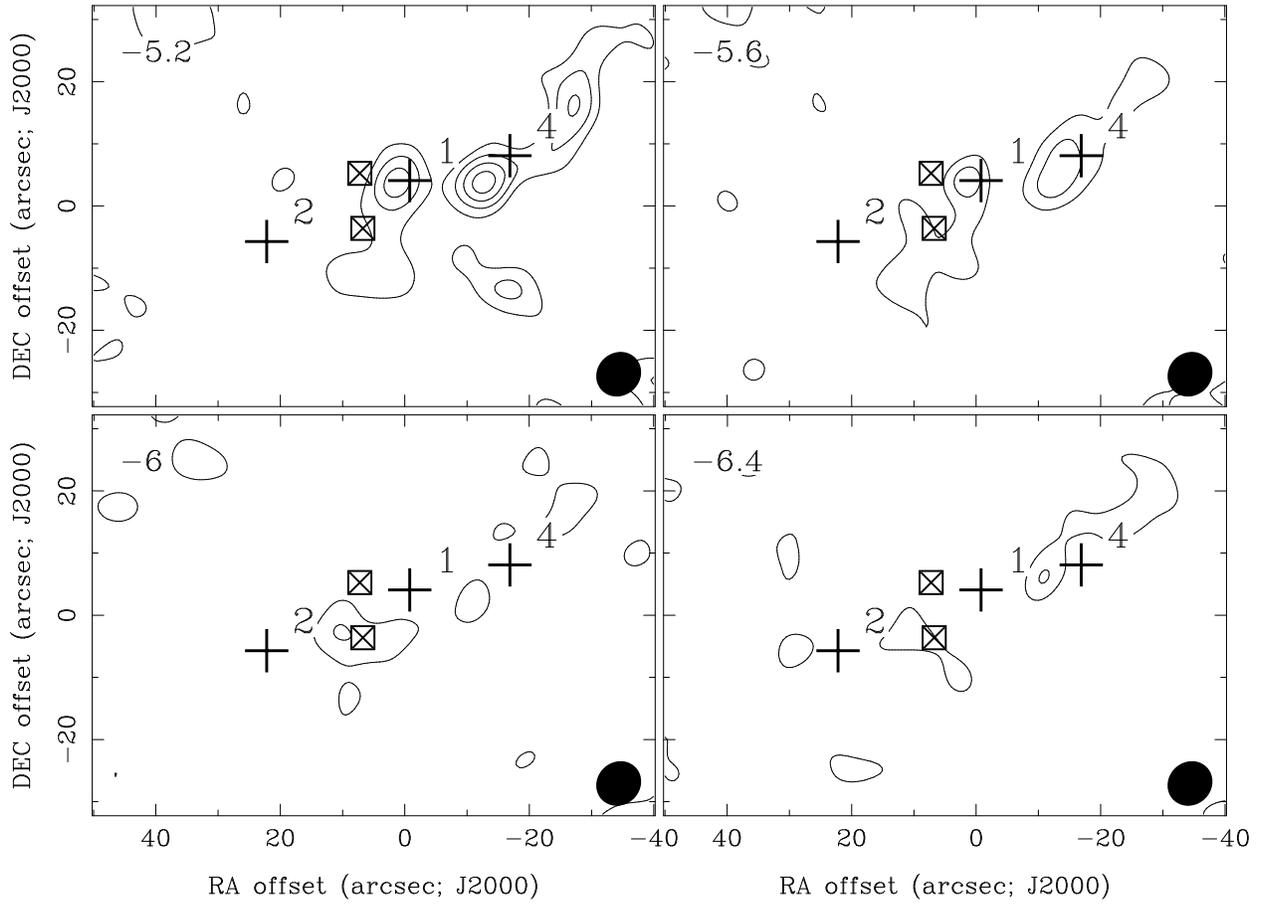}
\caption{
The channel map of blueshifted emission of HCO$^+$ 1$-$0 observed at the OVRO.
Symbols are the same as in Figure 14 and 15.
The contours begin at 2 $\sigma$ and increase in steps of 2 $\sigma$, where
1 $\sigma$ = 0.22 K.
}
\end{figure}

\clearpage

\begin{figure}
\figurenum{17}
\vskip -7cm
\plotone{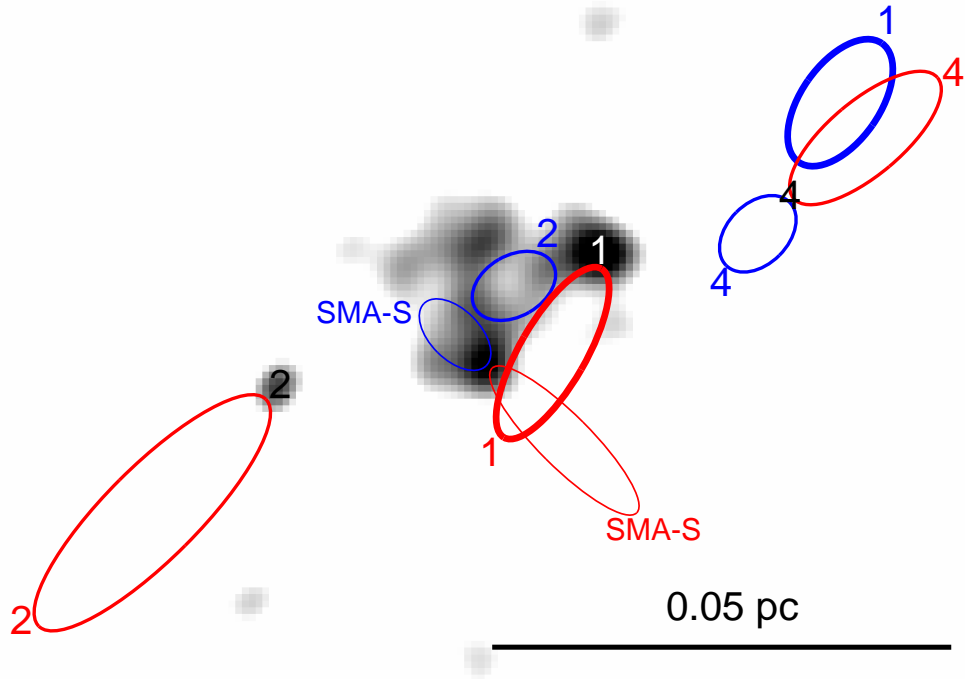}
\caption{
A possible explanation of the origin of each outflow. 
The grey image is the same as that in Figure 13.
Black and white numbers denote the positions of IRS1, IRS2, and IRS4. 
Red and blue lobes describe the red and blue components each outflow
identified in the channel maps of CO 2$-$1 and HCO$^+$ 1$-$0 observed at the
SMA and the OVRO, respectively.
Red and blue numbers and texts indicate the red- and blueshifted lobes 
originated from their corresponding sources.
}
\end{figure}

\clearpage

\begin{figure}
\figurenum{18}
\plotone{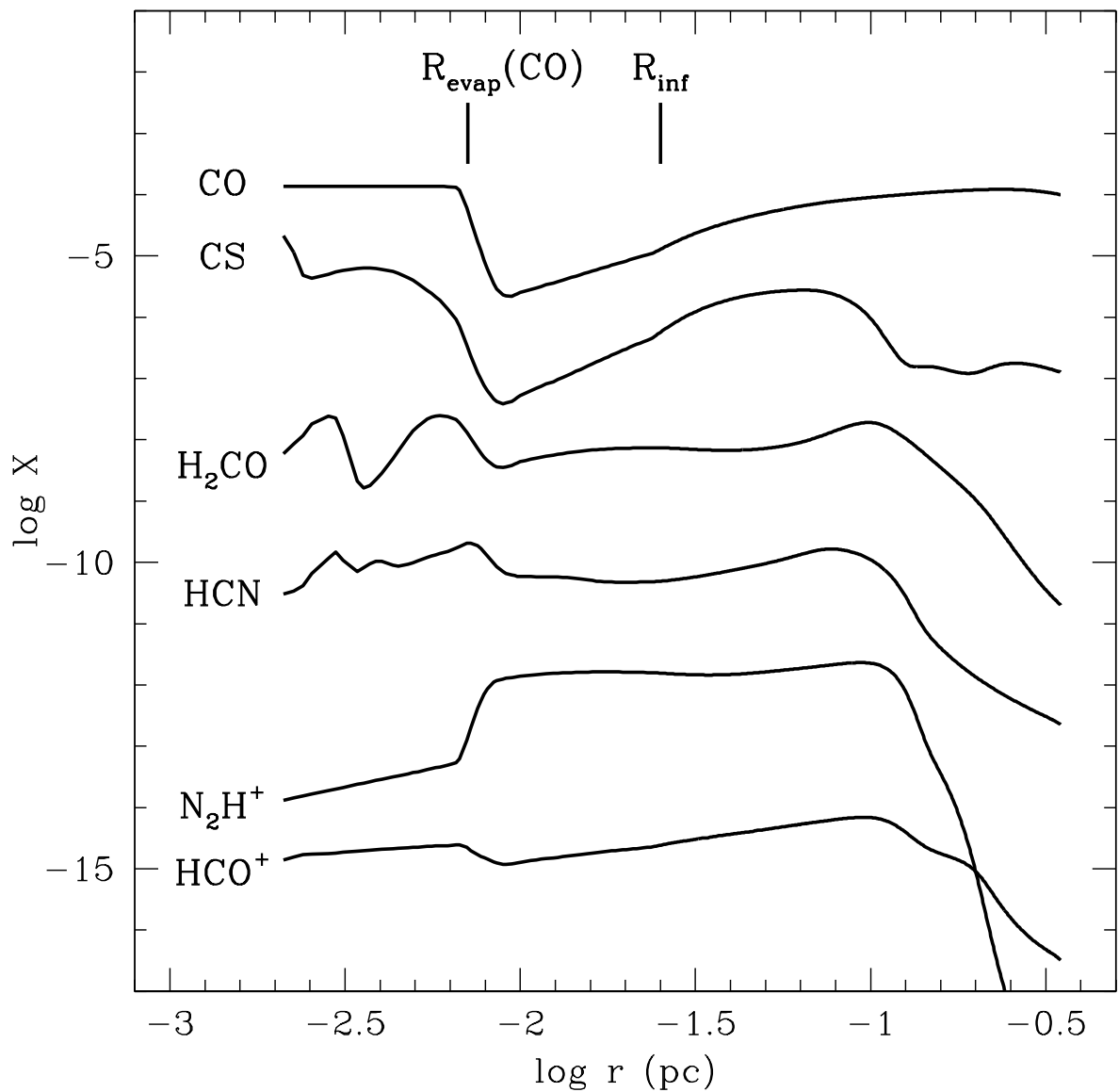}
\caption{
Abundance profiles of various molecules observed for this study. 
The profiles have been calculated 
using an evolutionary chemical model (Lee et al. 2004) defined on the basis 
of the physical parameters of L1251B.  Two vertical lines indicate the infall 
radius of the inside-out collapse and the radius where CO starts to evaporate.
Profiles are shifted up and down for easier comparisons, so the ordinate gives 
correct abundance of only CO.  Abundances of CS and H$_2$CO are shifted up by 
3.0 and 0.2 orders of magnitude respectively, while those of HCN, N$_2$H$^+$, 
and HCO$^+$ are shifted down by 1.0, 2.5, and 6.0 orders of magnitude 
respectively. 
}
\end{figure}

\end{document}